\documentclass[preprint,pre,showkeys,showpacs]{revtex4}
\newcommand{\beq}{\begin{equation}}
\newcommand{\eeq}{\end{equation}}
\usepackage{graphicx}
\usepackage{graphics}
\usepackage{latexsym}
\usepackage{subfigure}
\usepackage{multirow}
\usepackage{dcolumn}
\begin{document}

\title{Dynamic characterisers of spatiotemporal intermittency}

\author{Zahera Jabeen}
\email{zahera@physics.iitm.ac.in}
\author{Neelima Gupte}
\email{gupte@physics.iitm.ac.in}
\affiliation{Indian Institute of Technology-Madras, Chennai, India}
\date{\today}
\begin{abstract}

We study spatiotemporal intermittency in a system of coupled sine
circle maps. The phase diagram of the system shows parameter 
regimes where the STI lies in the directed percolation
class, as well as regimes which show pure spatial intermittency (where
the temporal behaviour is regular) which do not belong to the DP class.
Thus, both DP and non-DP behaviour can be seen in the same system.
The signature of DP and non-DP behaviour can be seen in the dynamic
characterisers, viz. the spectrum of eigenvalues of the linear
stability matrix of the evolution equation, as well as in the
multifractal spectrum of the eigenvalue distribution. The eigenvalue
spectrum of the system in the DP regimes is continuous, whereas it shows
evidence of level repulsion in the form of gaps in the spectrum
in the non-DP regime. The multifractal spectrum
of the eigenvalue distribution also shows the signature of DP and non-DP
behaviour. These results have implications for the manner in which
correlations build up in extended systems.

\end{abstract}
\maketitle

\section{Introduction}

The phenomenon of spatiotemporal intermittency (STI),
which is characterized by the coexistence of laminar states of regular
dynamics, and burst states of irregular dynamics, is ubiquitous in
natural and experimental systems.  Such behaviour has been seen in
experiments on convection \cite{cili,daviaud}, counterrotating Taylor-Couette flow \cite{colo},  oscillating ferro-fluidic spikes \cite{rupp}, and experimental and
numerical  studies of rheological fluids \cite{sriram, fielding}. In
theoretical studies, STI has been seen in PDEs such as the damped
Kuramoto-Sivashinsky equation \cite{kschate} and the one-dimensional  Ginzburg Landau equation \cite{glchate}, coupled map lattices \cite{kaneko} such as the Chat\'e-Manneville CML \cite{Chate}, the inhomogeneously coupled logistic map lattice \cite{ash}, and in cellular automata studies \cite{Chate}. \\
\indent A variety of scaling laws have been observed in these systems. However, there
are no definite conclusions about  their
universal behaviour. Many of the observed phenomena have been seen in experimental systems where no simple model is available.
There has been much discussion about the nature of spatiotemporal intermittency and its analogy with systems which undergo phase transitions.
It has  been argued that the transition to 
spatiotemporal intermittency with absorbing laminar states is a
second order phase transition, and that this transition falls in the same 
universality class as directed percolation \cite{Pomeau} with the
laminar states being identified with the `inactive' states and the turbulent states being identified as the `active' or percolating states. This conjecture has become the central issue in a long-standing debate
\cite{Rolf,Chate,Houlrik,Grassberger,Janaki}, which is still not completely resolved. Thus, the analysis of spatiotemporal intermittency remains a challenging theoretical problem.

In this paper, we study spatiotemporal intermittency in the coupled
sine circle map lattice \cite{Nandini}, a popular model for the
behaviour of mode-locked  oscillators. 
Spatiotemporal intermittency has been reported to exist for several points 
in the parameter space of this model and a full set of directed
percolation exponents has been found at these points \cite{Janaki, Zahera}. The detailed  
phase diagram of this model shows that these points lie on, or near,  the bifurcation boundary where the synchronized fixed points of the model lose stability. We now find that spatiotemporal intermittency can be found 
all along the bifurcation boundary of this region. Interestingly, while some points of this boundary show spatiotemporal intermittency where synchronized laminar regions coexist with turbulent regions, with associated directed percolation exponents (DP), other points of the boundary show pure spatial intermittency where the synchronized laminar regions are interspersed with bursts of temporally 
periodic or quasi-periodic behaviour which are not associated with DP exponents.Thus, both DP and non-DP regimes can be seen in this model.
The distinct signatures of these two types of behaviour can be found in  the
eigenvalue
distribution of the stability matrix calculated at one time step.
The eigenvalue
spectrum of the system in the DP regimes is continuous, whereas 
distinct gaps can be seen in the spectrum in the non-DP regime.
The
multifractal analysis of the eigenvalue distribution of the two cases also shows the signature of this behaviour. Thus, the signature of the DP and non-DP behaviour of the model can be found in the dynamic characterisers of the system. 
In the case of low dimensional systems, intermittency of different types  has been observed to contribute characteristic signatures to the distribution of finite time  Lyapunov exponents \cite{Ram}. The present study indicates the presence of a similar phenomenon in high-dimensional systems as well.   

\indent The organisation of this paper is as follows. The details of the model are given in section II. The phase diagram of this CML is discussed in the same section and the various types of STI observed are described therein. In section III, the universality classes are identified and the differences between the STI belonging to the DP class and non-DP classes  are quantified. Section IV describes dynamic characterisers which can pick up these distinct classes. The paper ends with a discussion of these results.

\section{The Model and the phase diagram}

The coupled sine circle map lattice has been known to model the
mode-locking behaviour \cite{gauri2} seen commonly in coupled
oscillators, Josephson Junction arrays, etc, and is also found to be
amenable to analytical studies \cite{Nandini}. The model is defined by 
the evolution equations
\beq
x_i^{t+1}=(1-\epsilon)f(x_i^t)+\frac{\epsilon}{2}[ f(x_{i-1}^t) + f(x_{i+1}^t) ]
\pmod{1}
\label{evol}
\eeq
	
where $i$ and $t$ are the discrete site and time indices respectively and $\epsilon$ is the strength of the coupling between the site $i$ and its two nearest neighbours. The local on-site map, $f(x)$ is the sine circle map defined as 
\beq
f(x)=x+\Omega-\frac{K}{2\pi}\sin(2\pi x)
\label{sine}
\eeq
Here, $K$ is the strength of the nonlinearity and $\Omega$ is the
winding number of the  single sine circle map in the absence of the
nonlinearity. We study the system with periodic boundary conditions in
the parameter regime  $0 < \Omega < \frac{1}{2\pi}$ (where
the single circle map has temporal period
1 solutions), $0 < \epsilon< 1$ and $K=1.0$. 
The phase diagram of the system is highly sensitive to initial
conditions due to the presence of many degrees of freedom and has been
studied extensively for several classes of initial conditions
\cite{Nandini, gauri2}, which result in  rich phase diagrams  with many distinct types  of attractors. In particular, this system 
has regimes of spatiotemporal intermittency (STI) when evolved
in parallel with random initial conditions \cite{Janaki}. 
An earlier study of the inhomogeneous logistic map lattice had shown that 
the bifurcation curves corresponding to bifurcations from the synchronized fixed point, 
can form rough guide-lines to the regions in parameter space where STI can be found \cite{ash}.  
It is therefore worthwhile
to investigate the detailed phase-diagram of the present system,
identify various types of dynamical behaviour, and correlate the
observed behaviour, especially the spatiotemporally intermittent behaviour,
with the known bifurcations that occur in the system.
\subsection{The Phase Diagram}
\begin{figure}[h]
\centering
\includegraphics[height=9cm,width=11cm]{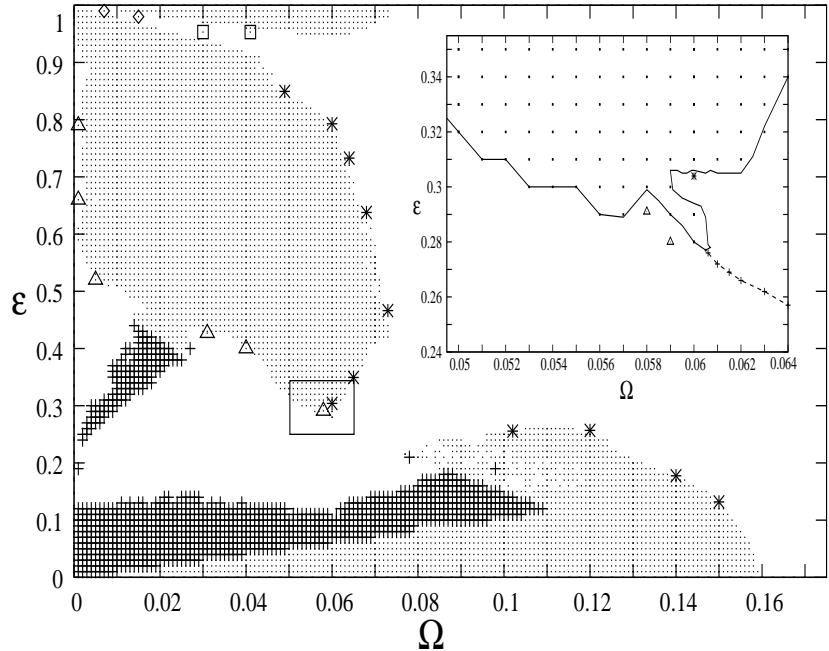}
\vspace{-.2in}
\caption{shows the phase diagram obtained at $K=1.0$ for a lattice of size $N =1000$. A transient of 15000 iterates has been discarded. The dots represent the  synchronized fixed point solutions. Plus ($+$) signs represent the cluster solutions. The diamonds
($\Diamond$) represent  STI with traveling (TW) wave laminar states and turbulent bursts whereas boxes ($\Box$) represent STI with TW laminar state and turbulent bursts containing solitons. The asterisks ($\ast$) represent STI belonging to the DP class, and spatial intermittency is represented by triangles ($\triangle$). The crossover regime from spatial intermittency to DP is magnified in the inset figure. SI with frozen bursts are seen below the dashed line. Above this line, the bursts start spreading on the lattice.\label{phd}}
\end{figure}
The phase diagram  of the system of Eqs. \ref{evol} and \ref{sine} for
the parameter region mentioned above is shown in Fig. \ref{phd}. Many
types of solution can be seen in the phase diagram. Stable,
synchronized, fixed point solutions, where the variables $x_i$  take
the value,
$x_i^t=\frac{1}{2\pi}\sin^{-1}(\frac{2\pi\Omega}{K})=x^{\star}$ for all
$i=1,\ldots N$ for all $t$, are indicated by dots.  These solutions,
which are very robust against perturbations, can be seen in large
regions of the phase diagram. Cluster solutions, in which $x_i^t=x_j^t$
for $i,j$ belonging to the same cluster are also seen in the phase
diagram, 
and are indicated  by $+$ signs in Fig. \ref{phd}.
The synchronized solution is, in fact, a single cluster solution where the size of the cluster is the lattice itself. 
Spatiotemporally intermittent solutions are seen near the bifurcation
boundaries where these synchronized solutions lose stability.
Several distinct kinds of STI are seen in the phase diagram. These are:
\begin{figure}
\begin{tabular}{cc}
(a)&(b)\\
{ 
\includegraphics[scale=.8]{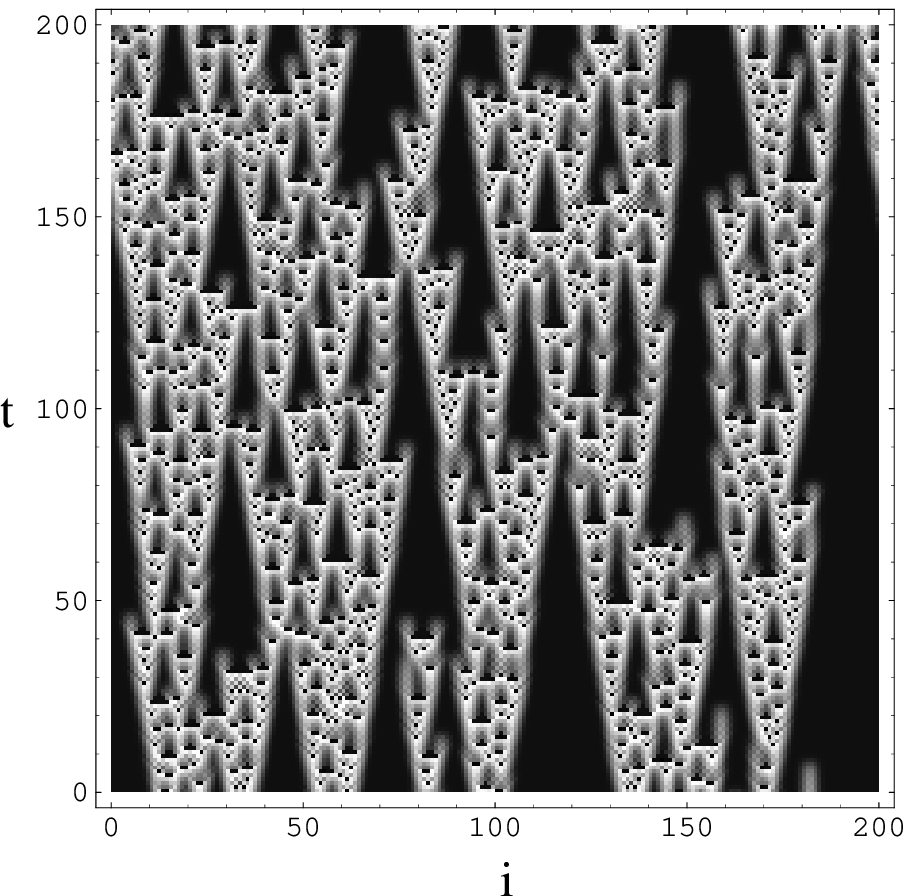}
}&
{\includegraphics[scale=.79]{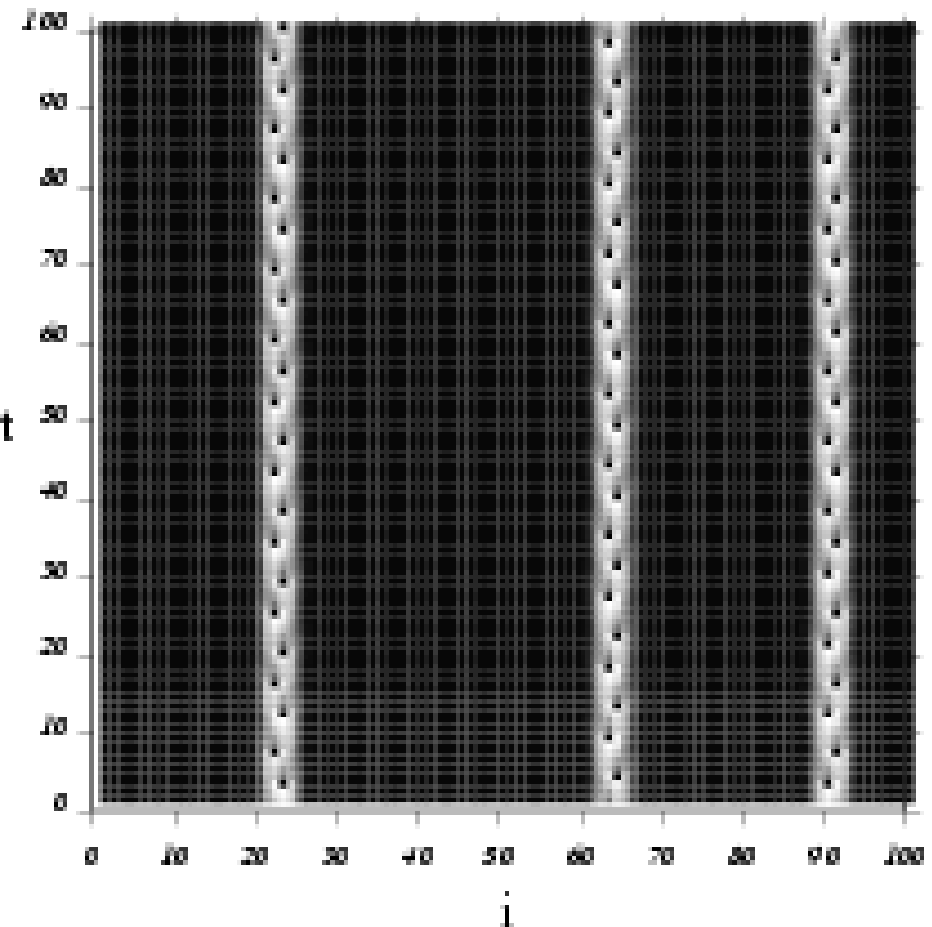}
}
\\
(c)&(d)\\
{
\includegraphics[scale=.8]{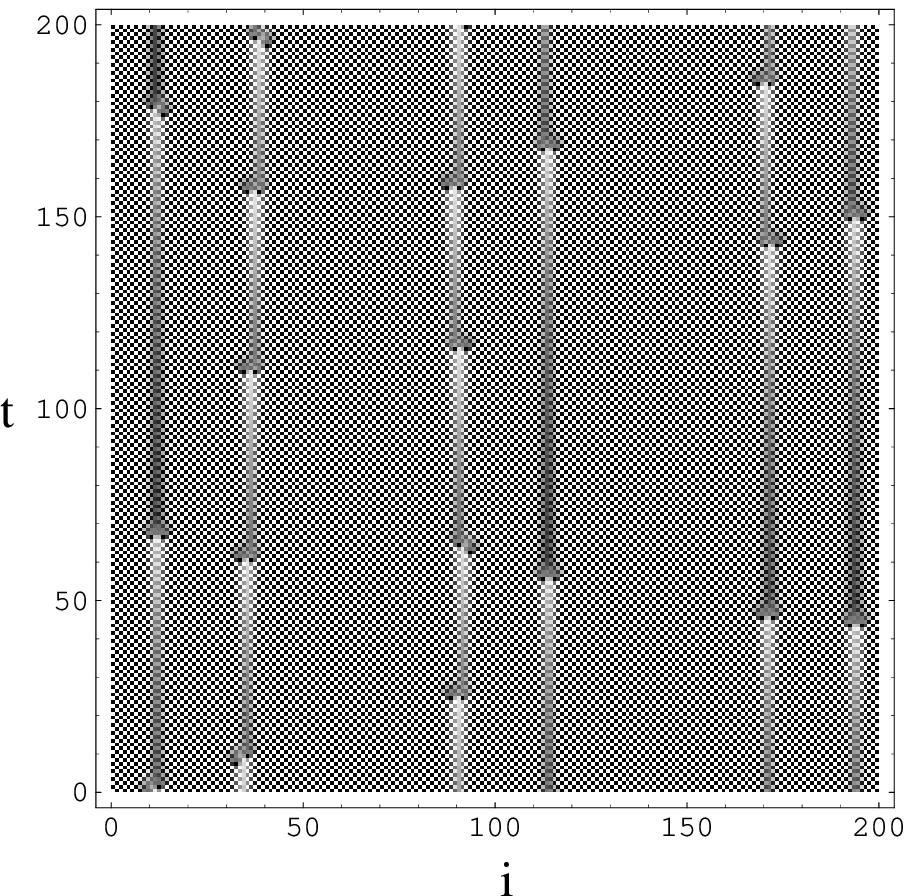}
}
&
{\includegraphics[scale=.8]{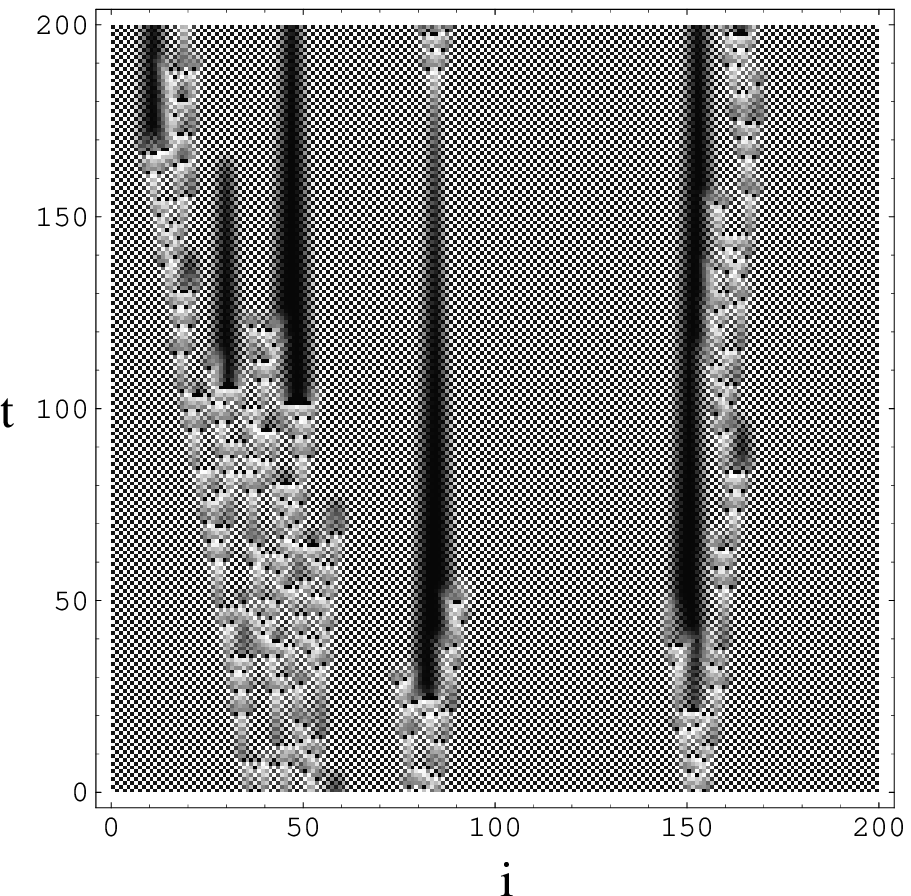}
}\\
\end{tabular}
\caption{ shows the space time plots of the different types of STI observed
in the phase diagram. The lattice index $i$ is along the x-axis and the
time index $t$ is along the y-axis. The space time plots show (a) STI with synchronized laminar state interspersed with turbulent bursts seen at $\Omega=0.06, \epsilon=0.7928$. (b) SI with synchronized laminar state with quasiperiodic and periodic bursts seen at $\Omega=0.031, \epsilon=0.42$. (c) STI with TW laminar state and turbulent bursts observed at $\Omega=0.007, \epsilon=0.99$. (d) STI with TW laminar state and turbulent bursts containing solitons at  $\Omega=0.03, \epsilon=0.954$. \label{stplot}}
\end{figure}
\begin{enumerate}
\item STI where the  synchronized laminar state is interspersed with
turbulent bursts is seen along the bifurcation boundary starting from
$\Omega =0.0457, \epsilon =0.89$ to $\Omega=0.06, \epsilon =0.30396$.
Some  of the points where this kind of STI is seen are shown by
asterisks ($\ast$) in the phase diagram. The laminar state corresponds
to the synchronized fixed point $x^{\star}$ defined earlier. The
turbulent state takes all values other than $x^{\star}$ in the [0,1]
interval. The space-time plot of these solutions is shown in Fig. \ref{stplot}(a).
\item Spatial intermittency (SI) with a synchronized laminar state
interspersed with quasi-periodic and periodic bursts is seen along the
boundary marked by triangles ($\triangle$) in the phase diagram. (The
triangles indicate specific locations where the SI has been studied).
The laminar state is the synchronized fixed point $x^{\star}$ and the
burst state is a mixture of quasi-periodic and periodic bursts. (See
Fig. \ref{stplot}(b)). 
\item  The locations where STI with travelling wave laminar states and
turbulent bursts can be seen are marked by diamonds ($\Diamond$) in the
phase diagram  (Fig. \ref{stplot}(c)). As can be seen from the space-time plot, the burst states are localized and do not spread through the lattice.
\item Parameter values which show spatiotemporal intermittency with
travelling wave laminar states and turbulent bursts are shown by boxes
($\Box$) in the phase diagram. The space-time plot of such states can be
seen  
in Fig. \ref{stplot}(d). These states differ from those seen in 
 Fig. \ref{stplot}(c) in that apart from the turbulent bursts, soliton-like structures which are turbulence of a coherent nature traveling in space and time, are also seen in this type of STI. Such coherent structures have also been seen in the Chat\'e-Manneville class of CMLs \cite{Grassberger, bohr}.
\end{enumerate}\vspace{-.1in}
We concentrate on STI with the simplest version of the laminar state, viz. the synchronized state. Two types of intermittent behaviour are associated with 
this laminar state, viz. spatiotemporal intermittency with spreading turbulent bursts, and spatial intermittency with localised periodic and quasi-periodic bursts. These two types of behaviour can be seen in contiguous regions of the boundary, but belong to different universality classes. We discuss this behaviour below, as well as the cross-over region between the two types of behaviour. STI with other types of laminar states, as well as STI in the presence of solitons
will be dealt with elsewhere.  The blank regions of the phase diagram
show solutions of varying degrees
of spatiotemporal irregularity which will also be discussed elsewhere.

\section{Universality classes in Spatiotemporal Intermittency}
We contrast the two types of intermittency and identify their universality classes in this section. It is seen that spatiotemporal intermittency with spreading bursts belongs to the directed percolation class, whereas spatial intermittency with periodic/quasi-periodic  bursts does not belong to the directed percolation class. The signature of this behaviour can be seen in the dynamic characterisers of the system, viz. the distribution of eigenvalues of the linear stability matrix.\vspace{-.2in}
\subsection{STI of the Directed Percolation class}
The phase diagram shows STI with synchronized laminar state interspersed with 
turbulent bursts 
along the upper boundary of the leaf shaped region where synchronized solutions are stable (the boundary on which asterisks are seen). These solutions show spreading and infective behaviour 
similar to that seen in  directed percolation models \cite{stauffer}. In
this type of STI, the spontaneous creation of turbulent bursts does not
take place as can be seen in the space-time plot of this STI (Fig.
\ref{stplot}(a)). A laminar site becomes turbulent only if it has been
infected by a neighbouring turbulent site at the previous time-step. The
turbulence either spreads to the whole lattice, or dies down completely
to the laminar state depending on the coupling strength, $\epsilon$.
Once all the sites in the lattice relax to the laminar state, it remains
in this state forever. Hence, the synchronized laminar state is the
absorbing state. Importantly, this type of STI, as seen in this model, is free of solitons which could bring in long-range correlations. Hence, a straightforward analogy with the DP class can be drawn in this case, where the burst states are identified with the 'wet' sites, the laminar states are the 'dry' sites and the time axis acts as the directed axis.
\begin{figure}
\begin{center}
\begin{tabular}{cc}
{\includegraphics[height=7.2cm,width=7.5cm]{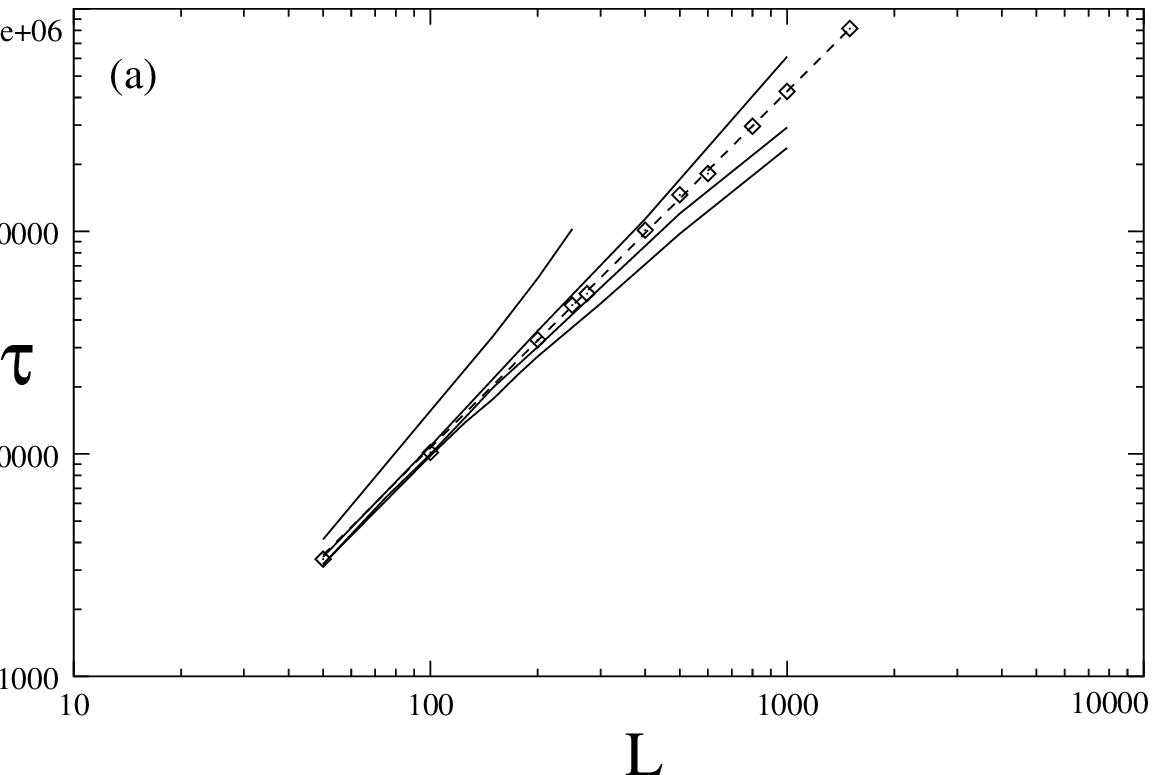}
}&

\hspace{0.5cm}\includegraphics[height=7.2cm,width=7.5cm]{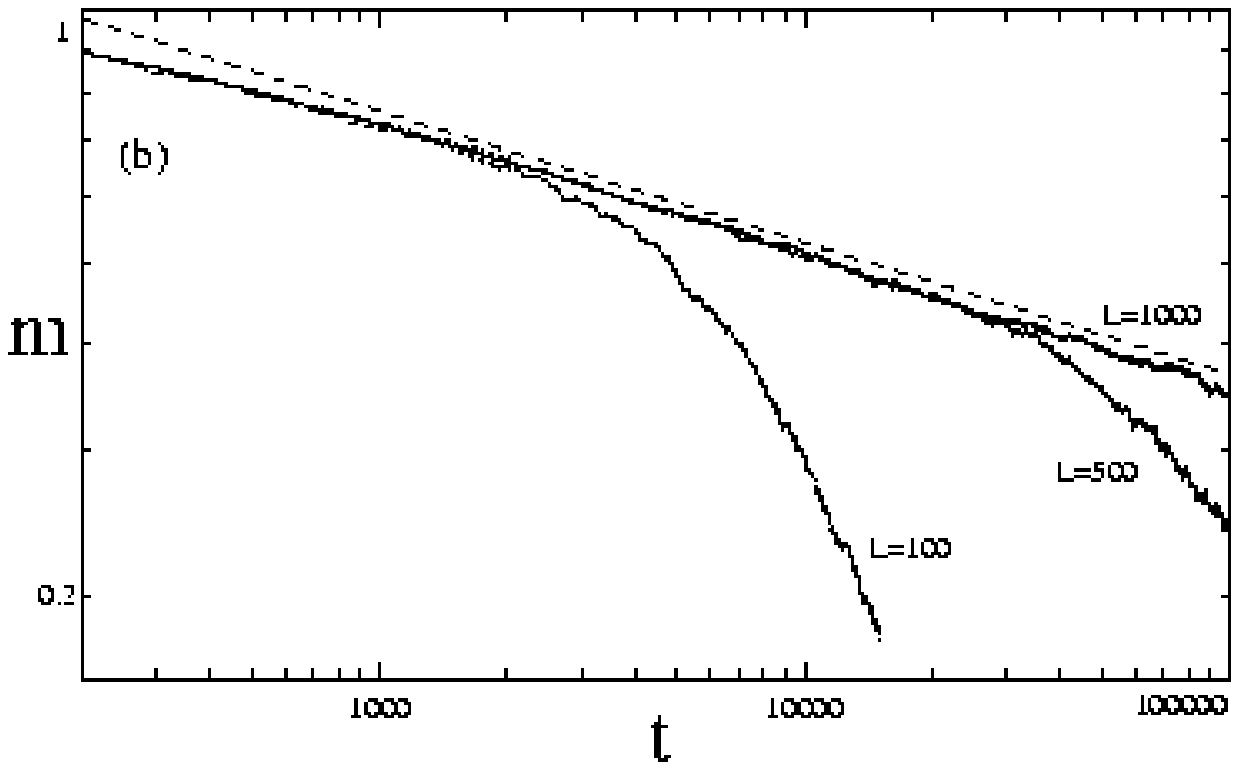}
\\ \\

\hspace{-.2cm}\includegraphics[height=7.5cm,width=7.9cm]{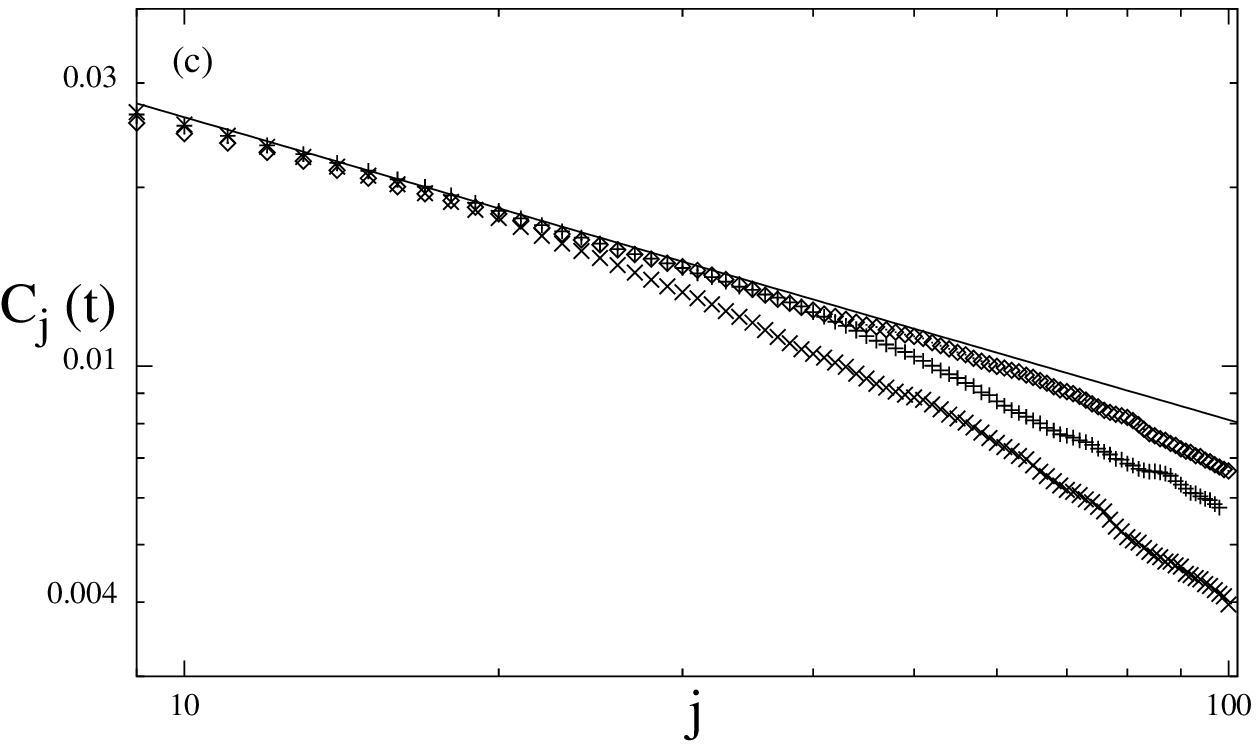}
&
{\hspace{.1cm}\includegraphics[height=7.5cm,width=7.9cm]{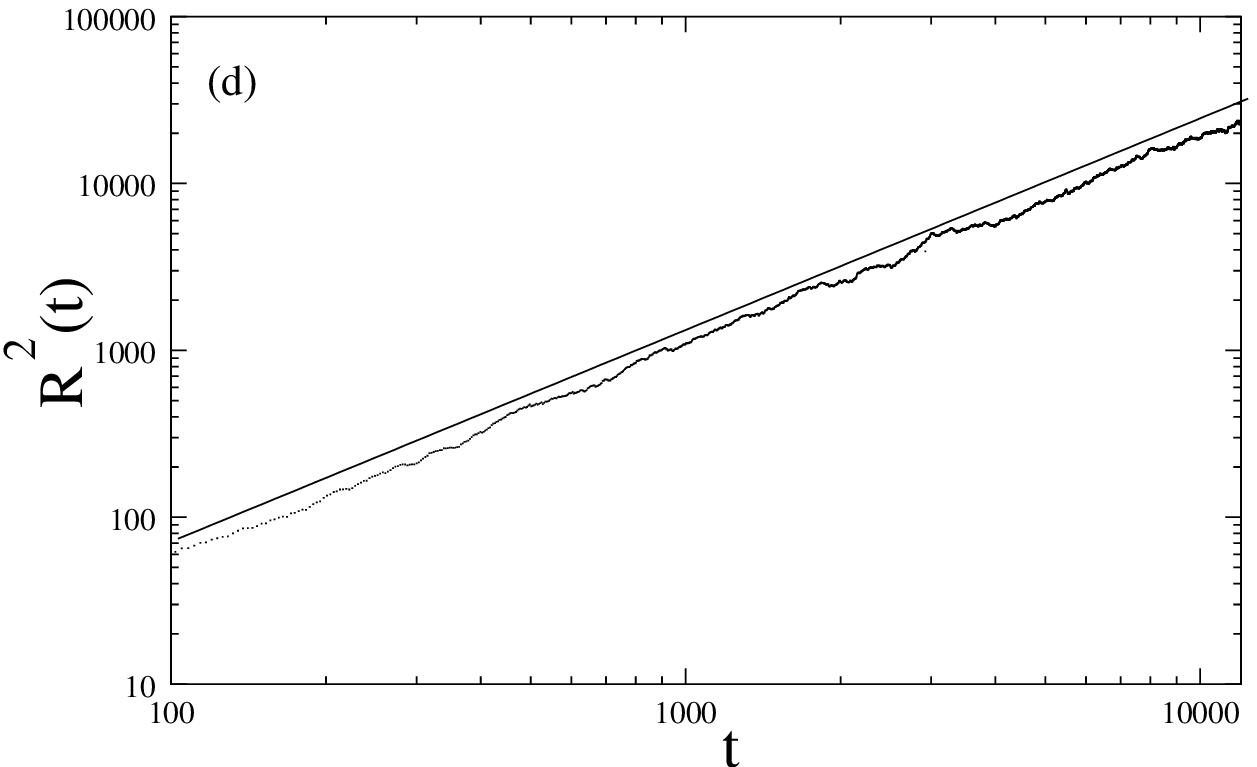}
}\\ \\

\end{tabular}
\caption{ (a) shows the log-log plot of the escape time, $\tau$ vs
lattice size $L$ at $\Omega=0.049$ and $\epsilon = 0.852, 0.8498,
0.8495, 0.8492, 0.849$ from top to bottom. At $\epsilon_c = 0.8495$,
$\tau$ scales with $L$ with $z=1.60\pm0.01$. (b) shows the log-log plot
of order parameter, $m$ vs $t$ at $\Omega=0.073,\epsilon=0.4664$. The
exponent $\beta/{\nu z} = 0.157$. (c) shows the log-log plot of the
correlation function $C_j(t)$ vs $j$ at $\Omega=0.049,\epsilon=0.8495$
at timesteps $t=65000, 35000, 15000$ shown as $\Diamond, +, \times$
respectively. The exponent $\eta'-1=0.51$ (d) shows the log-log plot of
the radius of gyration, $R^2(t)$ vs $t$ at
$\Omega=0.073,\epsilon=0.4664$. $z_s=1.268$. All logarithms are to base 10.
 \label{dpplots}}

\end{center}
\end{figure}

The DP transition is characterised by a set of static  critical exponents associated with  physical quantities of interest such as the escape time, the order parameter which is defined as the fraction of turbulent sites in the lattice at time $t$, the distribution of laminar lengths, and  the pair correlation function. In addition, a set of dynamic critical exponents can be obtained by considering 
temporal evolution from initial conditions which
correspond to an absorbing background with a localised disturbance, i.e.
a few contiguous sites which are different from an absorbing background.
The quantities of interest are, the time dependence of $N(t)$,
the number of active sites at time $t$ averaged over all runs, $P(t)$,
the survival probability, or the fraction of initial conditions which
show a non-zero number of active sites (or a propagating disturbance)
at time $t$ and the radius of gyration $R^2(t)$, which is defined as the mean squared deviation of position of the active sites from the original sites of the turbulent activity, averaged over the surviving runs alone.
The detailed definition of the full set of DP exponents is given in the
appendix and typical behaviour is shown in Fig. \ref{dpplots}. This
complete set of DP exponents has been calculated at the points marked by
asterisks in the phase diagram (Fig. \ref{phd}). (The exponents at two of these points viz. at parameters $\Omega=0.068, \epsilon=0.63775$, $K=1.0$, and $\Omega=0.064, \epsilon=0.73277$, $K=1.0$                                    were found in an earlier work \cite{Janaki}). 
\begin{table}
\begin{center}
\begin{tabular}{|l|l|c|c|c|c|c|c|c|}
\hline
\multirow{2}{*}{ $\Omega$}&\multirow{2}{*}{ $\epsilon_c(\Omega)$}&\multicolumn{7}{c|}{ Bulk exponents}\\
\cline{3-9}
 & &  $z$ &  $\beta/{\nu z} $ &  $\beta$ &  $\nu$ &  $\eta'$ & { $\zeta$} & {$\zeta'$}\\
\hline
 	&	&	&	&	&	&	&	&\\
  0.049	 &0.8495&  1.60$\pm$.01 & 0.16$\pm$0.01 &  0.271 &  1.1&  1.51$\pm$0.0 & 1.68$\pm$0.01	&0.87$\pm$0.01 \\
  0.06	&0.7928	&1.59$\pm$0.02	&0.17$\pm$0.02	&0.293	&1.1	&1.51$\pm$0.01	&1.68$\pm$0.01	&0.78$\pm$0.01	\\
  0.073	&0.4664	&1.58$\pm$0.02	&0.16$\pm$0.01	&0.273	&1.1	&1.5$\pm$0.01	&1.65$\pm$0.01	&0.72$\pm$0.01	\\	
 0.065	&0.34949	&1.59$\pm$0.03	&0.16$\pm$0.01	&0.273	&1.1	&1.5$\pm$0.01	&1.66$\pm$0.01	&0.75$\pm$0.01	\\
 0.06	&0.30396	&1.6$\pm$0.02	&0.16$\pm$0.01	&0.27	&1.05	&1.5$\pm$0.01	&1.61$\pm$0.01	&0.70$\pm$0.01	\\
 0.102	&0.25554	&1.6$\pm$0.01	&0.16$\pm$0.00	&0.277	&1.1	&1.52$\pm$0.01	&1.67$\pm$0.01	&0.73$\pm$0.01	\\
 0.12	&0.257	&1.60$\pm$0.01	&0.15$\pm$0.01	&0.264	&1.1	&1.51$\pm$0.01	&1.64$\pm$0.01	&0.71$\pm$0.01	\\ 
\hline
\multicolumn{2}{|c|}{\bf DP}& {\bf 1.58}	& {\bf 0.16}	& {\bf 0.28}	& {\bf 1.1}	& {\bf 1.51}	&{\bf 1.67}	&{\bf 0.748}	\\
	
 \hline
\end{tabular}
\vspace{0.2in}
\caption{ The static exponents obtained  at the critical $\epsilon_c$ are shown in the above table. The universal DP exponents are listed in the last row. These exponents have been obtained after averaging over 1000 initial conditions. \label{dpt}}
\end{center}
\end{table}
\nopagebreak
\begin{table}
\begin{center}
\begin{tabular}{|c|c|c|c|c|}
\hline
\multirow{2}{*}{$\Omega$}	&\multirow{2}{*}{$\epsilon_c$}	&\multicolumn{3}{c|}{Spreading Exponents}\\
\cline{3-5}
	&	&$\eta$	&$\delta$	&$z_s$	\\
\hline
	&	&	&	&	\\

0.049	&0.8495	&0.308$\pm$0.002	&0.17$\pm$0.02	&1.26$\pm$0.01	\\
0.06	&0.7928	&0.315$\pm$0.007	&0.16$\pm$0.01	&1.26$\pm$0.01\\
0.073	&0.4664	&0.308$\pm$0.001	&0.17$\pm$0.01	&1.27$\pm$0.00\\
0.065	&0.34949	&0.303$\pm$0.001	&0.16$\pm$0.01	&1.27$\pm$0.01\\
0.06	&0.30396	&0.317$\pm$0.001	&0.17$\pm$0.01	&1.26$\pm$0.0\\
0.102	&0.25554	&0.315$\pm$0.001&0.16$\pm$0.00	&1.25$\pm$0.01\\
0.12	& 0.257		&0.305$\pm$0.00	&0.16$\pm$0.00	&1.27$\pm$0.03\\
\hline
\multicolumn{2}{|c|}{\bf DP}	&{\bf 0.313}	&{\bf 0.16}	&{\bf 1.26}	\\
\hline	
\end{tabular}
\end{center}
\caption{The spreading exponents obtained at $\epsilon_c$ are shown in the above table. The last row lists the DP exponents. \label{dpst}}
\end{table}

All these points are located near the bifurcation boundary of the spatiotemporally synchronized solutions. The static and dynamic exponents obtained after averaging over $10^3$ initial conditions at these parameter values have been listed in Table \ref{dpt} and \ref{dpst} respectively. The agreement between these exponents and the universal DP exponents is complete.

\subsection{Spatial intermittency shows non-DP behaviour}

Spatial intermittency  with synchronized laminar state, and
quasi-periodic or periodic bursts, is also seen in the vicinity of the
bifurcation boundary at the locations indicated by triangles. The
laminar state is the synchronized fixed point $x^\star$ defined earlier.
The bursts observed are quasi-periodic in nature.  Fig. \ref{stplot}(b) shows the space-time plot of SI. We can see the absence of spreading dynamics or infective behaviour on the lattice. The bursts are spatially localized which is unlike the dynamics seen in directed percolation systems. 
The spatially intermittent solutions have zero velocity components in the spatial direction, and modes which travel along the lattice do not appear. In addition to the solutions with quasiperiodic bursts seen in the space-time plot, solutions which have strictly periodic bursts can also be seen.

\begin{figure}
\begin{center}
\includegraphics[height=7.0cm,width=16cm]{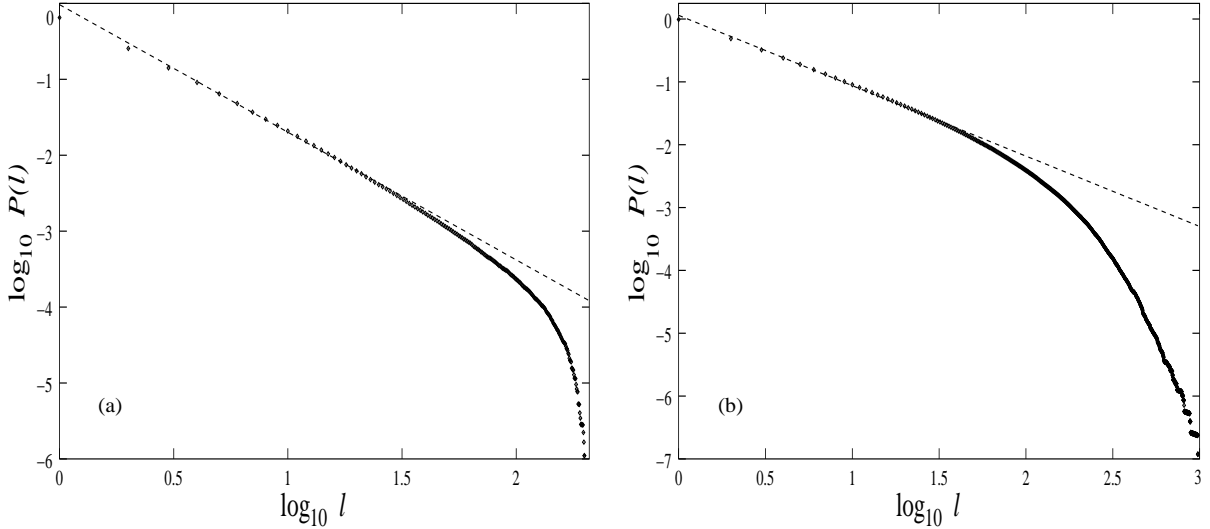}
\caption{ shows the $\log-\log$ (base 10) plot of the laminar length distribution for (a) STI with synchronized laminar state obtained at $\Omega=0.06, \epsilon=0.7928$. The exponent obtained is 1.681. (b) SI obtained at $\Omega=0.04, \epsilon=0.4$. The exponent $\zeta$ is 1.12. \label{lam}}

\end{center}
\end{figure}
The scaling exponent $\zeta$ for the laminar length distribution
$P(\ell)\sim \ell^{-\zeta}$, is found to have the value  $ 1.1$ in the
case of SI which is very different from the corresponding DP exponent
($\zeta_{DP}=1.67$) (Fig. \ref{lam}). Hence, SI does not belong to the DP universality class. We note that this exponent, however, has been seen for the inhomogenously coupled logistic map lattice \cite{ash} where similar spatial intermittency is seen.

\subsection{Cross-over Regime}

\begin{figure}
\begin{center}
\begin{tabular}{cc}
{(a)} & {(c)} \\
{\includegraphics[scale=.85]{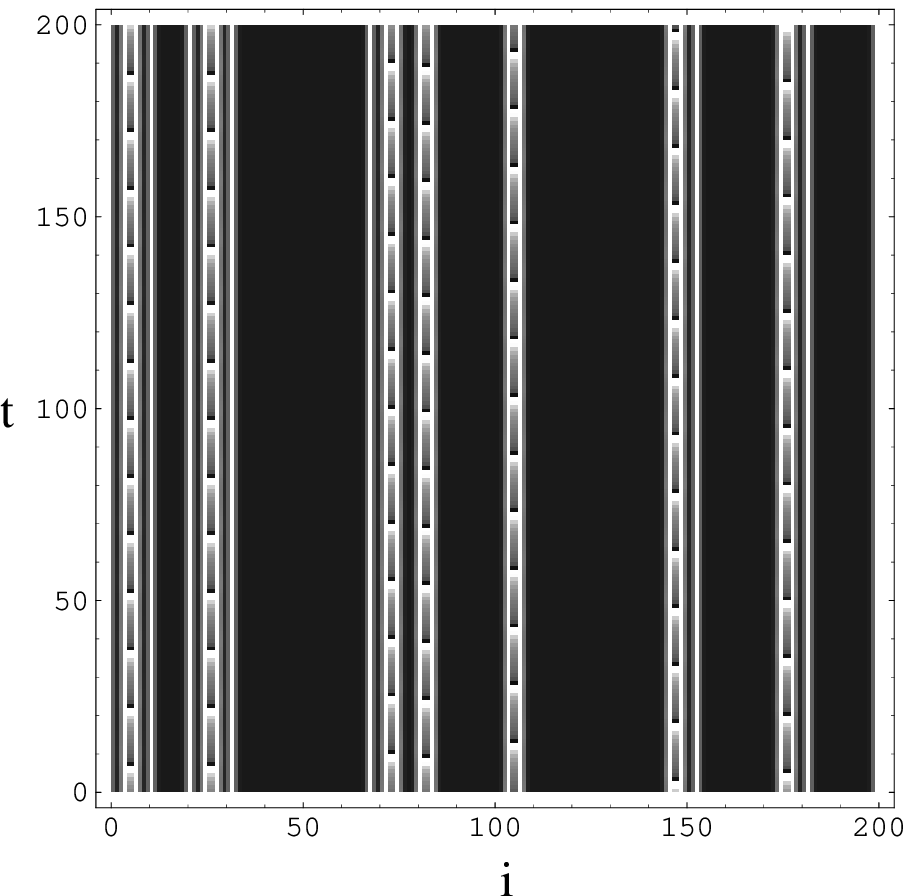}\vspace{-0in}} &
{\includegraphics[scale=.85]{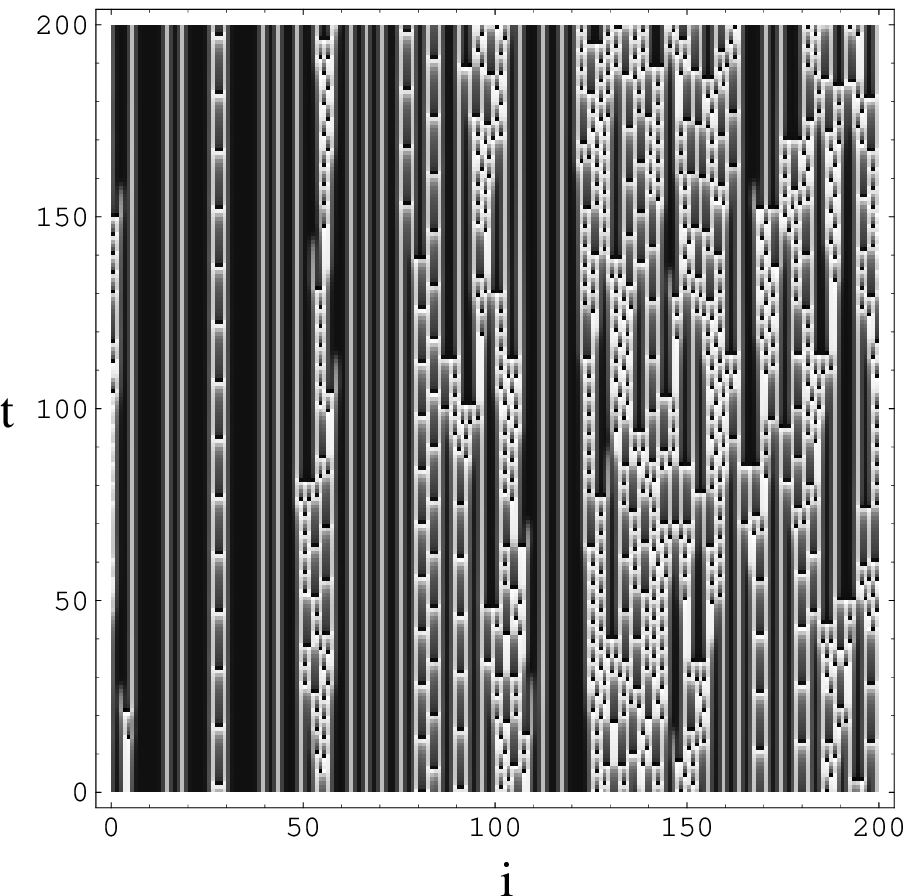}\vspace{-0in}} \\
{(b)} & {(d)}\\
 \hspace{-.0in} { \includegraphics[height=7.2cm,width=7.5cm]{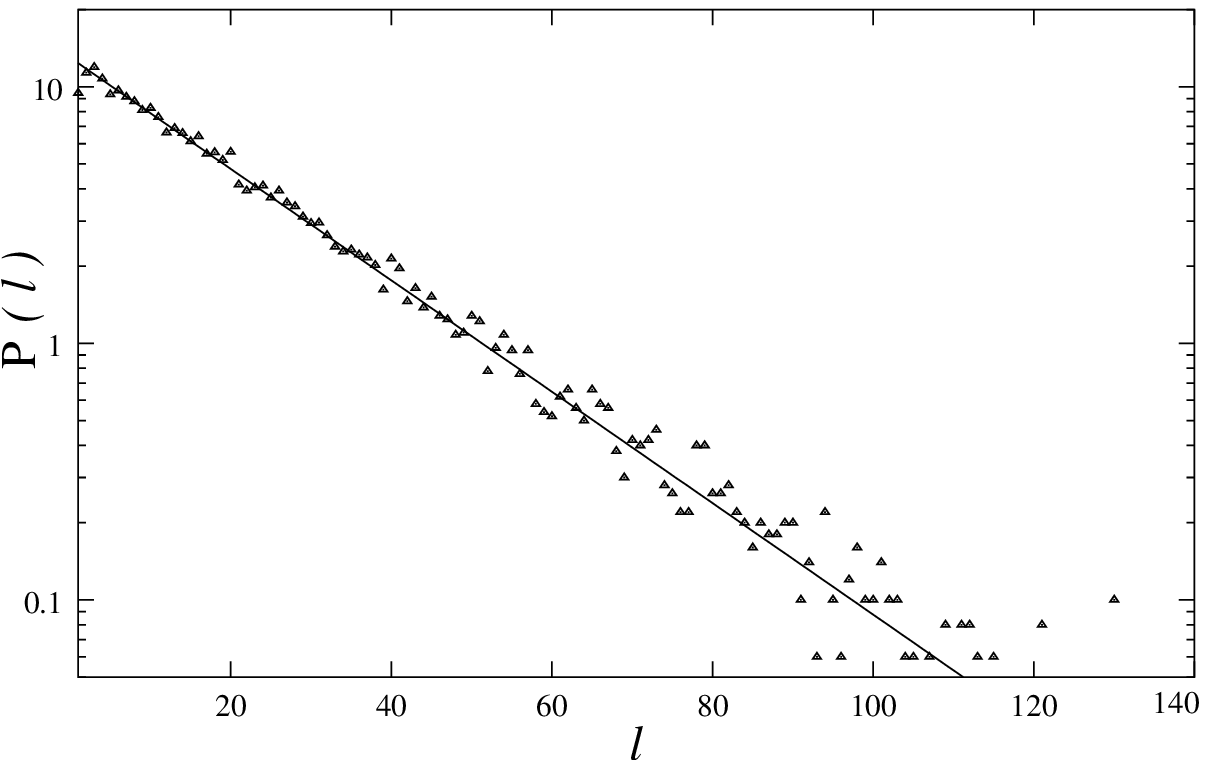} } &
\hspace{-.0in}{\includegraphics[height=7.2cm,width=7.8cm]{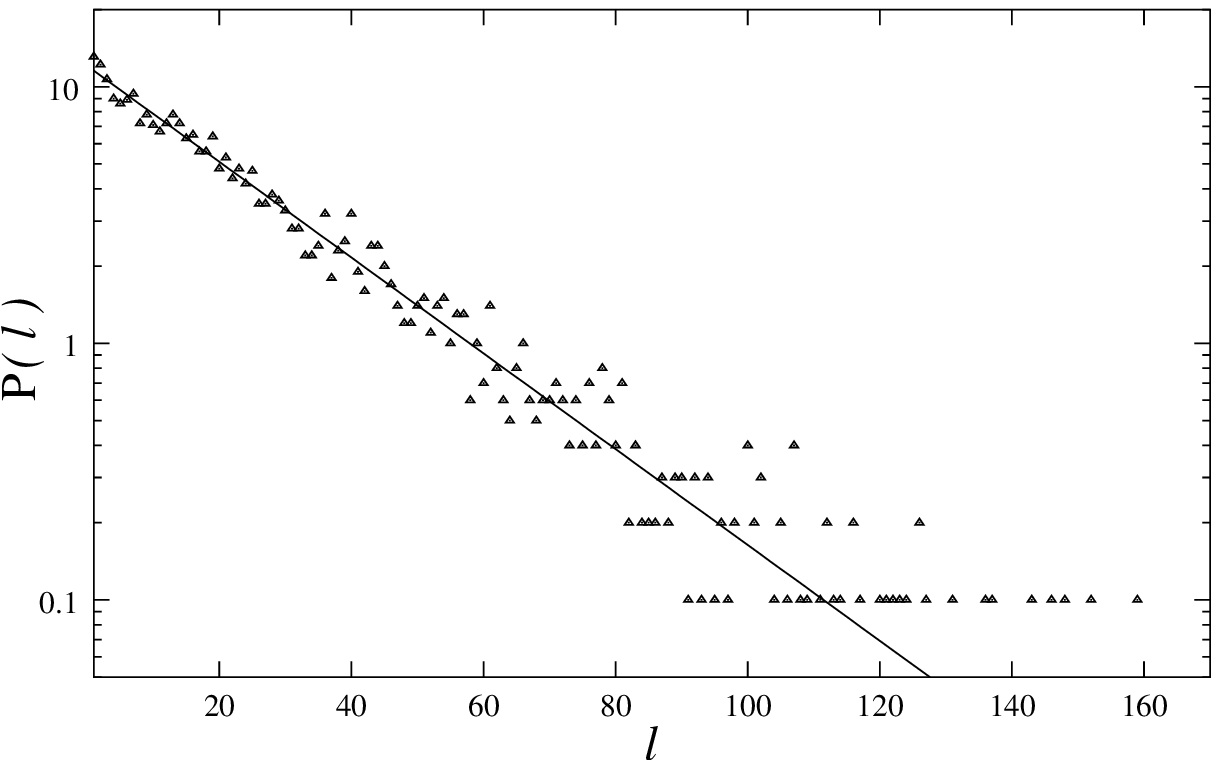}} \\

\end{tabular}
\caption{ shows the behaviour on either side of the crossover line shown
in Fig. \ref{phd} (inset). Fig 5(a) shows the space-time plot of SI seen at $\Omega=0.0608, \epsilon=0.273$. Frozen bursts can be seen in the space-time plot. The laminar length distribution  is plotted on a semi-log plot (base 10) in Fig 5(b) . The fit to this distribution is $13.0~ \exp~ (-0.05 x)$  Fig 5(c) shows the  space-time plot obtained at $\Omega=0.061, \epsilon=0.273$. The laminar length distribution is shown in Fig 5(d). The fit is $12.06~ \exp~(-0.043 x) $ \label{exp}}
\end{center}
\end{figure}

It is clear from the phase diagram that the regimes of spatial intermittency 
and spatiotemporal intermittency are contiguous to each other on the
lower part of the bifurcation boundary of the synchronized solutions (
in the neighbourhood of  $\Omega=0.06$ and $\epsilon=0.3$ ), and cross-over effects can be expected in this parameter regime.   
This cross-over region is magnified in the inset to Fig. \ref{phd}. The
transition takes place through an intermediate stage wherein apart from
periodic and quasi-periodic bursts, frozen bursts are also seen. This
regime is found just below the dashed line shown in the inset, e.g. at
$\Omega=0.0608$, $\epsilon=0.273$. Here, the laminar length distribution
is exponential in nature (See Fig. \ref{exp}). Above the dashed line (e.g. at $\Omega=0.061$ and $\epsilon=0.273$), the crossover starts with the appearance of bursts which spread on the lattice. As $\epsilon$ is increased further, the bursts lose their localised nature and STI with synchronized laminar state is seen. At the point marked by an asterisk ($\ast$) in the figure, DP exponents are obtained.

The distinction between the DP and non-DP regimes lies in the extent to which 
the burst solutions are able to spread into the laminar regions, i.e.   
the extent to which they are able to infect the laminar regions. While the spreading exponents are the obvious signature to this problem, their non-universal 
nature, and strong dependence on the initial configuration makes their use problematic. However, the dynamic signature of the extent to which burst solutions can spread and mix into the laminar regions, is contained in the spectrum of the eigenvalue   
distribution of the one-step stability matrix and also in the multifractal spectrum of the eigenvalue distribution. The eigenvalue spectrum of the DP class
is continuous whereas the non-DP class contains distinct gaps in the spectrum 
indicating regions where the eigenvalues are repelled, corresponding to stretching rates which are excluded. 
These gaps appear to lead to the strong spatial localisation and temporally regular/quasi-regular behaviour for the burst solutions characteristic of spatial intermittency.
The multifractal spectrum of the eigenvalues also contains the signature of this behaviour. Thus, the eigenvalue spectrum and the
multifractal spectrum of the eigenvalues, constitute dynamic characterisers 
of spatiotemporal intermittency. We elaborate on these characterisers in the next section.

\section{Dynamic signatures of DP and non-DP class}

\subsection{The eigenvalue distribution of the stability matrix}
The linear stability matrix of the evolution equation \ref{evol} at one time-step about the  solution of interest is given  by the  $N \times N$ dimensional matrix, $M_t^N$, given below
\begin{displaymath}
\mathbf{M_t^N} = 
\left( \begin{array}{ccccc}
\epsilon_s f'(x_1^t) & \epsilon_n f'(x_2^t) & 0 & \ldots& \epsilon_n f'(x_N^t)\\
\epsilon_n f'(x_1^t)&\epsilon_s f'(x_2^t)&\epsilon_n f'(x_3^t)&\ldots&0\\
0&\epsilon_n f'(x_2^t)&\epsilon_s f'(x_3^t)&\epsilon_n f'(x_4^t)&\ldots\\
\vdots&\vdots&\vdots&\vdots&\vdots\\
\epsilon_n f'(x_1^t)&0&\ldots&\epsilon_n f'(x_{N-1}^t)&\epsilon_s f'(x_N^t)\\ 
\end{array}\right)
\end{displaymath}

where, $\epsilon_s=1-\epsilon$, $\epsilon_n=\epsilon/2$, and  $f'(x_i^t)=1-K\cos (2\pi x_i^t)$. $x_i^t$ is the state variable at site $i$ at time $t$, and a lattice of $N$ sites is considered. 

The diagonalisation of $M^N_t$ gives the $N$ eigenvalues of the stability matrix.  
 The eigenvalues of the stability matrix were calculated for spatiotemporally intermittent solutions which result from bifurcations from the spatiotemporally synchronized solutions. 
The eigenvalue distribution for the STI belonging to the DP universality class and SI was calculated by averaging over 50 initial conditions. 

\begin{figure}[ht]
\begin{center}
\begin{tabular}{c}
\hspace{-.4in}\includegraphics[height=7.5cm,width=17.1cm]{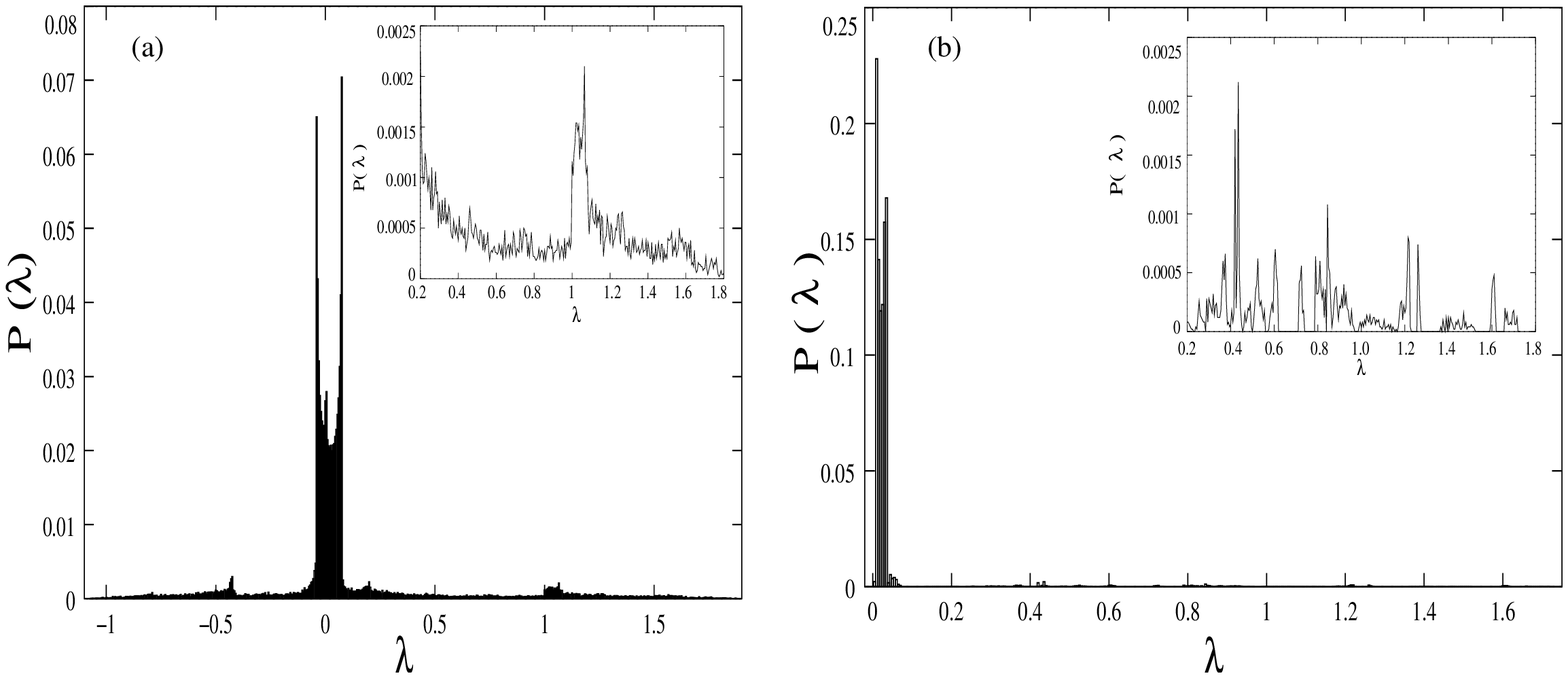}\\
\hspace{-.4in}\includegraphics[height=7.5cm,width=17.1cm]{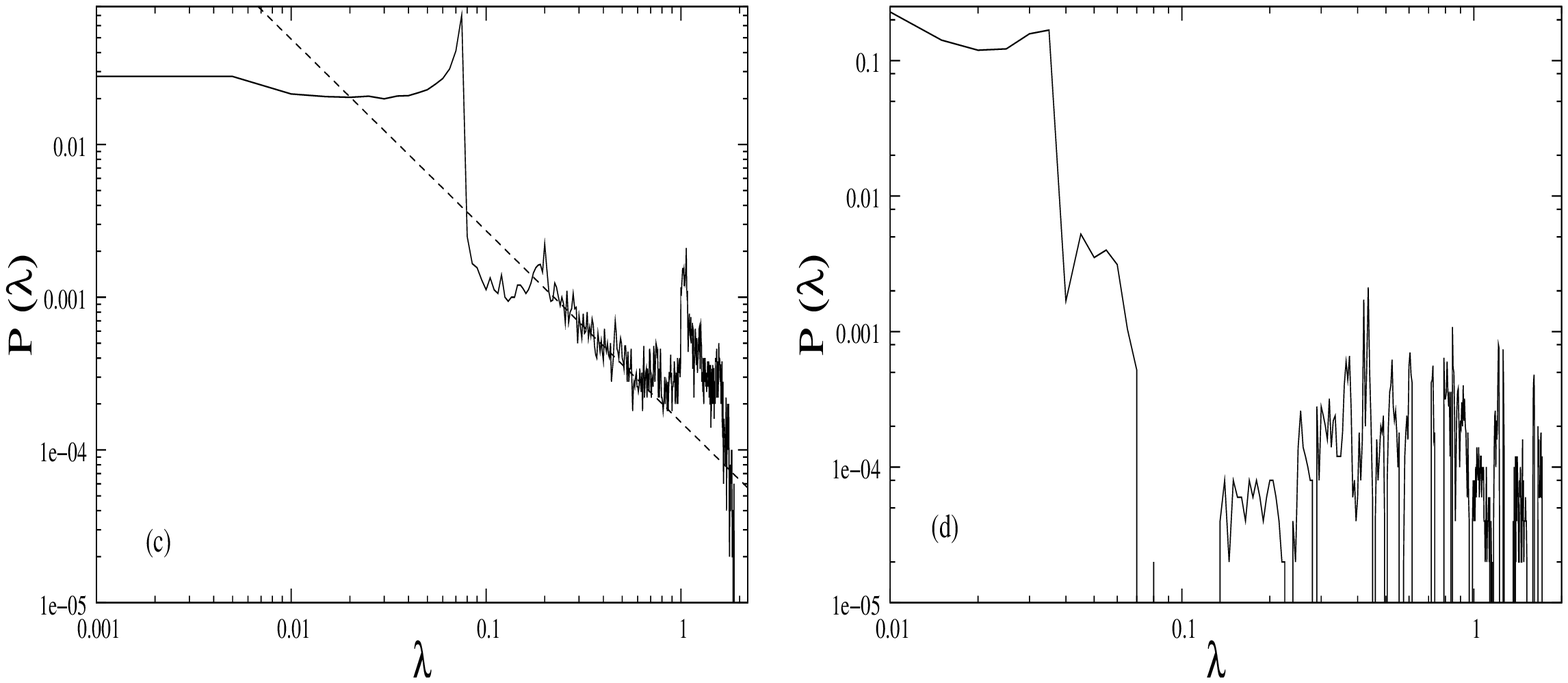}
\end{tabular}
\caption{ shows the eigenvalue distribution for (a) STI belonging to
the DP class at $\Omega=0.06,\epsilon=0.7928$ and (b) Spatial
intermittency at $\Omega=0.04,\epsilon=0.4$. A section of the
eigenvalue distribution is magnified in the inset figures. Gaps are
seen in the spatial intermittency eigenvalue distribution whereas the
eigenvalue distribution for STI does not show any such
gaps.\label{nhgm} The logarithmic  plots (base 10) of $P(\lambda)$
versus $\lambda$ can be
seen in Fig. 6(c) for the DP class (for the positive part of the
spectrum) and Fig. 6(d) for the spatial
intermittency.}
\end{center}
\vspace{-.2in}
\end{figure}

The eigenvalue distributions for STI belonging to the DP class can be seen in the Fig. \ref{nhgm}(a), and  that for spatial intermittency can be seen 
in Fig. \ref{nhgm}(b). 
It is clear from the insets,  the eigenvalue spectrum of the SI case shows distinct gaps. No such gaps are seen in the eigenvalue spectrum of the STI belonging to the DP class and the spectrum is continuous. Thus, a form of level repulsion is seen in the eigenvalue distribution for parameter values which show spatial intermittency. 
We note that such gaps are seen at all the parameter values studied
where spatial intermittency is seen, and that no gaps are seen for any
of the parameter values where DP is seen \cite{FN}. It can be seen that
\ref{nhgm}(c) shows power-law scaling (with power $-1.254$) in the range 
$0.1$ to $1$ unlike Fig. \ref{nhgm}(d). The gaps in the spectrum can
also be seen in Fig. \ref{nhgm}(d).
Thus, the gaps in the spectrum are associated with 
temporally quasi-periodic or periodic bursts in a synchronized fixed
point laminar background. The  bursts have no velocity component along
the lattice, and hence do not travel in space, nor do they infect
their laminar neighbours. On the other hand, when the spectrum is continuous the bursts  are temporally turbulent and show the infective behaviour characteristic of 
\pagebreak
directed percolation problems.
Since the eigenvalue spectrum of the SI case has zero probability regions, they are strongly picked up by multifractal analysis. However, it is interesting to note that the multifractal analysis of this case, with gaps excluded, also carries the signature of spatial intermittency.  
We discuss these signatures in the next section.
\begin{figure}[ht]
\begin{center}
\begin{tabular}{cc}
\hspace{-.5in}{\includegraphics[height=7.05cm,width=8cm]{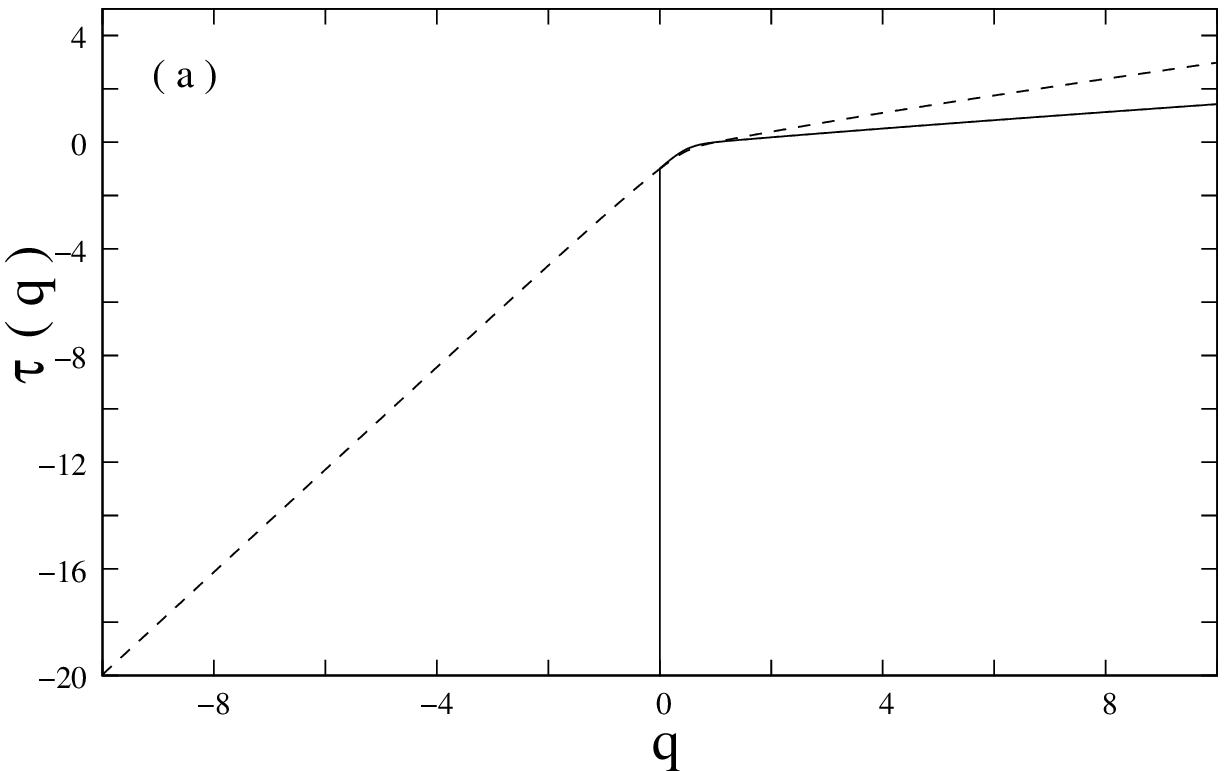}}&
{\hspace{.2in}\includegraphics[height=7.1cm,width=8.2cm]{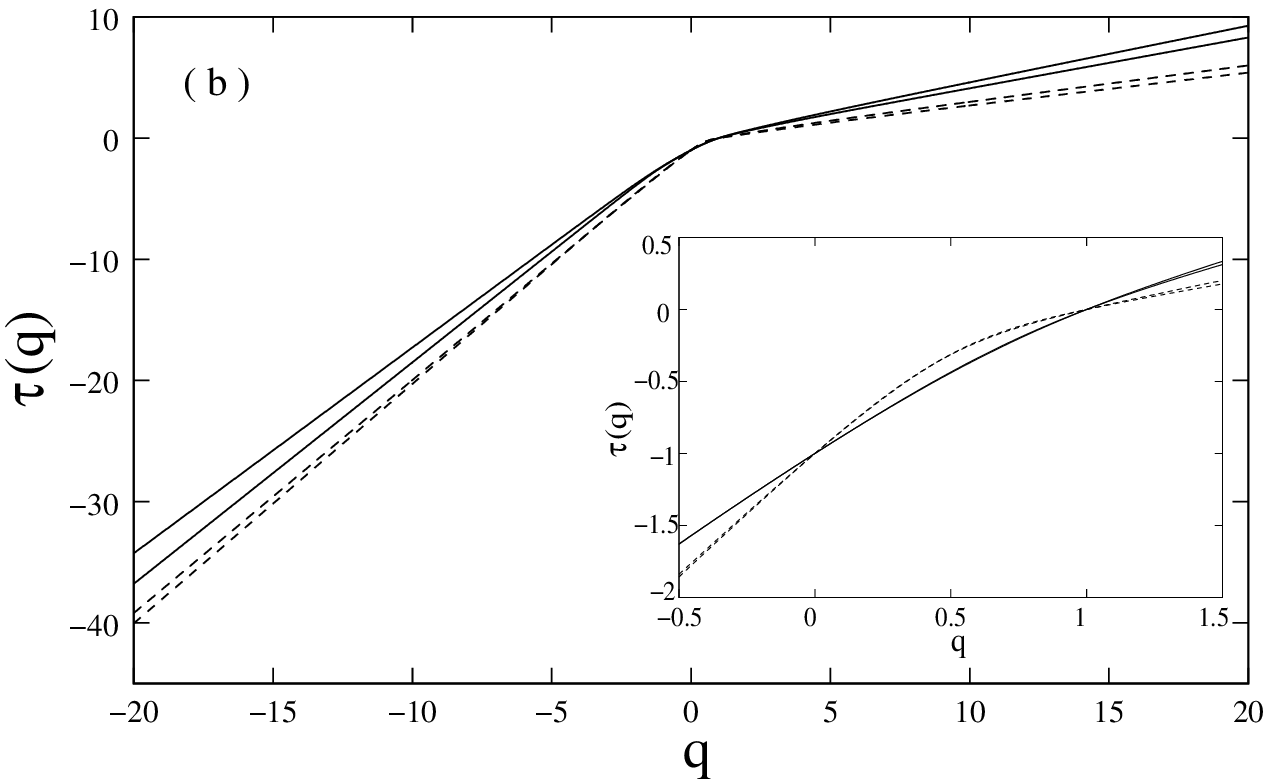}}\\
\end{tabular}
\caption{(a) shows the $\tau(q)$ vs $q$ curves for $\Omega=0.058,\epsilon=0.291$ including the gaps (solid line) and excluding the gaps (dashed line); (b) shows the $\tau (q)$ vs $q$ curves (with gaps excluded) for $\Omega=0.073,\epsilon=0.4664$; $\Omega=0.06,\epsilon=0.30396$; $\Omega=0.058,\epsilon=0.291$ and $\Omega=0.04,\epsilon=0.4$ from top to bottom. The first two curves represent STI belonging to the DP class (shown by solid lines) and the last two represent spatial intermittency. The difference in the curvatures of $\tau(q)$ for the DP and the non-DP class can be seen in the inset figure in which the region near $q=0.0$ has been magnified. \label{tauq}}
\end{center}
\end{figure}

\subsection{ Multifractal Analysis}
The eigenvalue distributions obtained at different parameter values were analysed using the multifractal framework \cite{halsey, rang}. 
Given a probability distribution whose support is covered by equal
lengths, the generalized dimensions, $D_q$, are
defined by the relation
\beq
D_q = \frac{1}{q-1} ~ \lim_{l \rightarrow 0} ~~\frac{\log \sum_{i} p_i^q}{\log l}
\eeq  
where $p_i$ is the probability associated with the $i^{th}$ bin and $l$
is the bin-size. Clearly, $D_q$  picks out the effect of larger probabilities at large positive $q$'s and smaller probabilities at large negative $q$'s. The quantity $\tau (q)$ is defined as 
\beq
\tau (q)= D_q (q-1)  
\eeq
The $\tau (q)$ vs $q$ spectrum for the STI and SI cases are shown in Fig. \ref{tauq}. Fig. \ref{tauq}(a) has been plotted for the parameter values 
 $\Omega=0.058,\epsilon=0.291$, where spatial intermittency is seen, and there are gaps in the spectrum. The solid line is the plot of $\tau$ versus $q$ for the case where the entire support of the distribution is covered with equal lengths of $s=0.005$. It is clear that $\tau(q)$ diverges to $-\infty$ for negative values of $q$, due to the presence of gaps in the spectrum where the distribution takes zero values. 

\begin{center}
\begin{figure}[h]
\begin{center}
\begin{tabular}{c}
{\hspace{-.8in}\includegraphics[height=8cm,width=17cm]{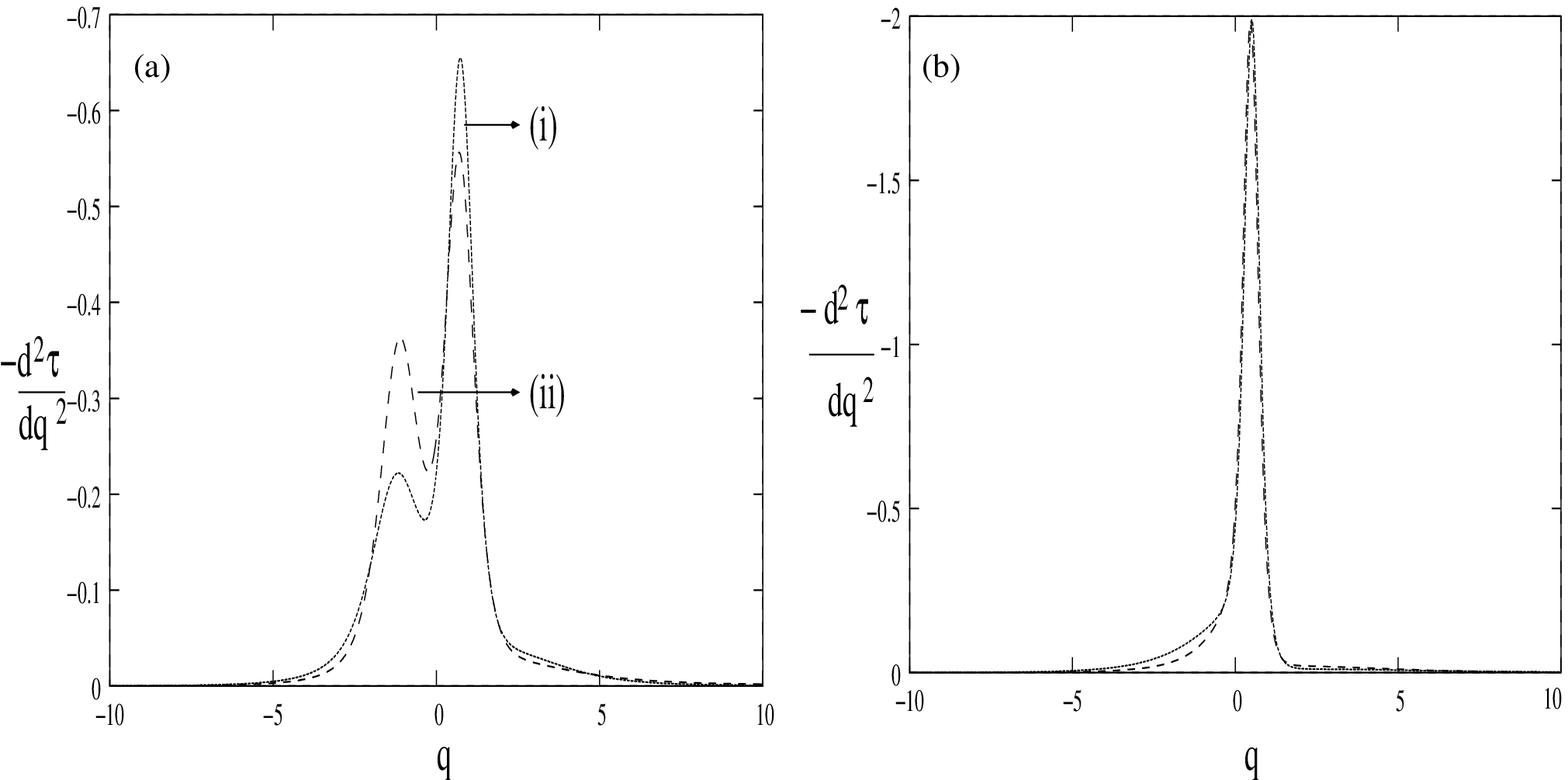}}\\
{\includegraphics[height=8cm,width=9.5cm]{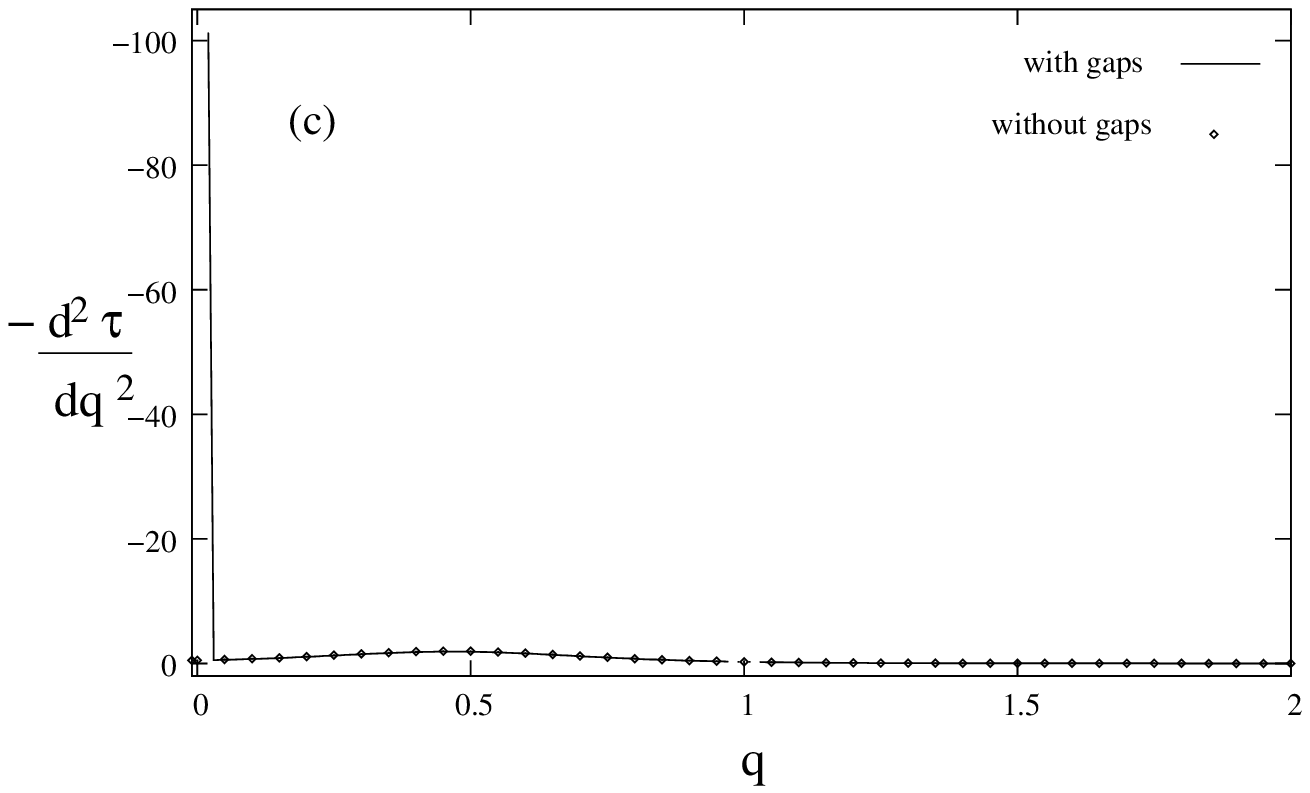}}\\
\end{tabular}
\caption{ shows $-\frac{d^2 \tau (q)}{dq^2}$ plotted against $q$ for (a) STI belonging to DP class calculated at (i) $\Omega=0.073, \epsilon=0.4664$ and (ii) $\Omega=0.06, \epsilon=0.30396$; and for (b) spatial intermittency at parameters (i) $\Omega=0.058, \epsilon=0.291$ and (ii) $\Omega=0.031, \epsilon=0.42$. Twin peaks are seen for STI belonging to the DP class; (c) shows $-\frac{d^2 \tau (q)}{dq^2}$ obtained for SI including gaps and excluding gaps. A sharp change is seen near $q=0.0$ when gaps are included.\label{curva}}
\end{center}
\end{figure}
\end{center}

\begin{center}
\begin{figure}[!hp]
\begin{center}
\begin{tabular}{c}
{\hspace{-.8in}\includegraphics[height=7cm,width=17.1cm]{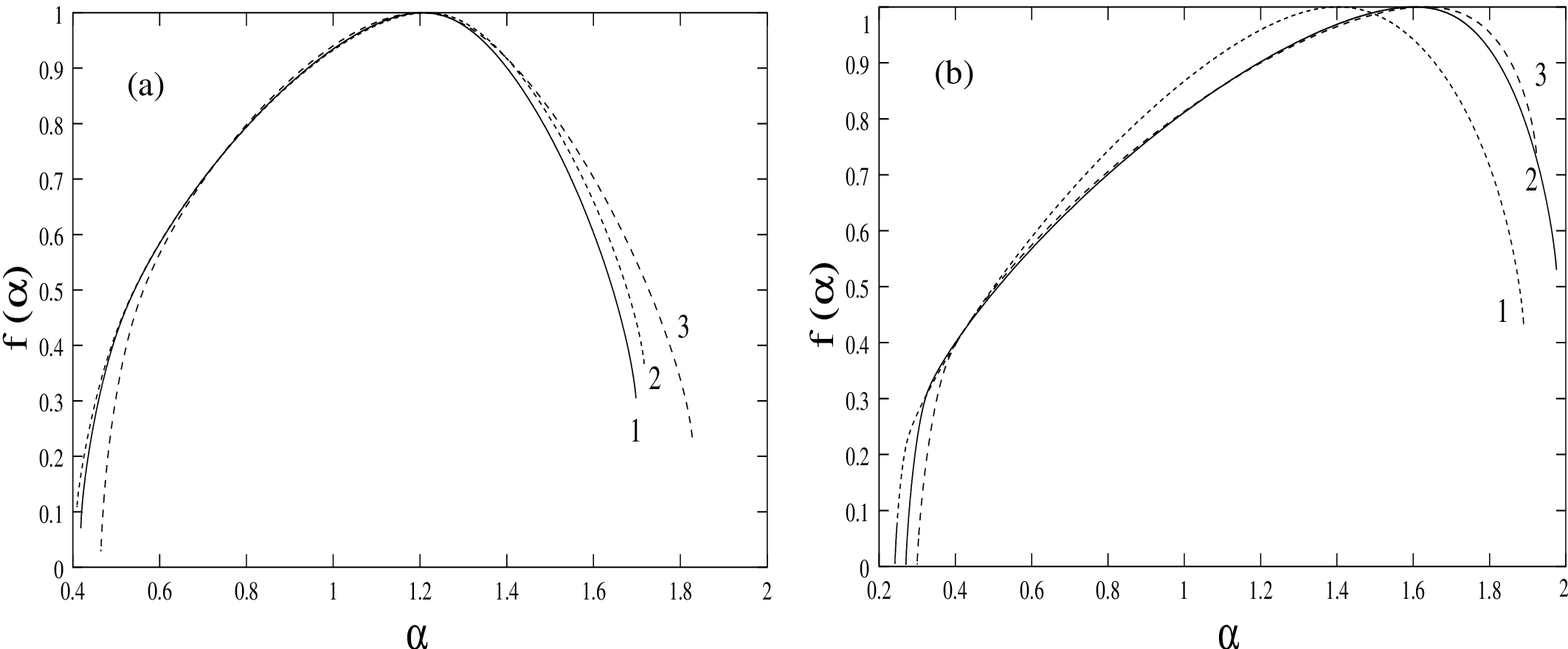}}\\ \\
{\includegraphics[height=7.2cm,width=8.7cm]{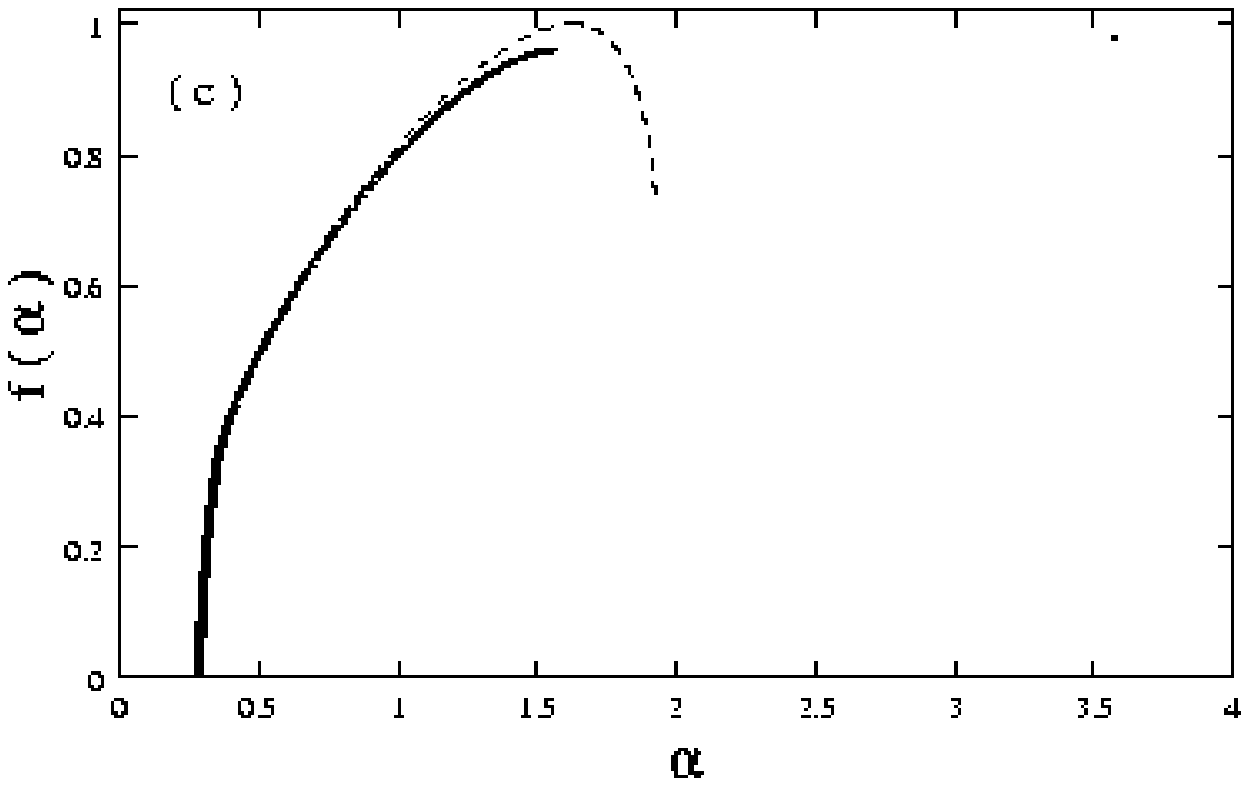}}\\
\end{tabular}
\caption{ shows the $f(\alpha)$ vs $\alpha$ curves for (a) STI belonging
to the DP class at parameters ($\Omega=0.06,\epsilon=0.7928$),
($\Omega=0.064, \epsilon=0.73277$), ( $\Omega=0.073, \epsilon=0.4664$)
labelled as 1,2,3 respectively in Fig. (a) and; (b) Spatial
Intermittency at parameters ($\Omega=0.031,\epsilon=0.42$), (
$\Omega=0.04, \epsilon=0.4$) and ($\Omega=0.058, \epsilon=.291$)
labelled as 1, 2, and 3  respectively in Fig.(b). The $f(\alpha)$
spectrum for the DP class (Fig.(a) overlap with each other for small
positive $q$'s whereas the $f(\alpha)$ curves for SI class (Fig.(b)) do not coincide with those of the DP class; (c) shows the $f(\alpha)$ curve obtained with gaps ($\Diamond$) and without gaps (line).\label{falph}}
\end{center}
\end{figure}
\end{center}

The dotted line in 
 Fig. \ref{tauq}(a)
corresponds to the $\tau$ versus $q$ curve
obtained for the same distribution without including  the contribution of the gaps. While the $\tau(q)$ now no longer diverges, its behaviour is still distinct from that obtained from  the distributions which correspond to DP regimes. This can be seen in Fig. \ref{tauq}(b).
We see that the $\tau(q)$ curves show  different behaviour in the neighbourhood of $q=0$ for the DP(solid lines) and non-DP (dotted lines) cases. 
The behaviour in the vicinity of the knee of the curve is magnified and shown in the inset. The dotted lines corresponding to the SI case fall on the same curve here, and are distinct from the curve on which the solid lines of the DP case fall (although both sets of curves separate out for large values of $|q|$). It is clear that the curvature of the $\tau$ versus $q$ curve in this region is different near $q=0$ for the DP and non-DP cases. 
The curvature of $\tau(q)$, $\frac{d^2\tau(q)}{dq^2}$ for the two types
of STI is shown in Fig. \ref{curva}. The negative and positive parts
of the y-axis have been interchanged  for
ease of representation. A twin peak is seen in the curvature of $\tau(q)$ of STI belonging to the DP class (Fig. \ref{curva}(a)) whereas a single peak is seen in the case of spatial intermittency (Fig.\ref{curva}(b)). Fig. \ref{curva}(c) shows the curvature of the SI case, for $q$ positive, with gaps included (solid line), and gaps excluded ($\Diamond$). A jump is seen in the spectrum at $q=0.0$ due to the contribution of the gaps, and the two curvatures coincide
completely for $q$ positive.\\
\indent The signatures of the $DP$ versus non-DP behaviour can also be seen in the $f-\alpha$ curves of the distribution. 
The $\alpha(q)$, the scaling exponent of the probabilities, and
$f(\alpha)$, the fractal dimension of the set which supports the
probability which scales with the exponent $\alpha$, are obtained from the relations  
$\alpha(q)=\frac{d\tau(q)}{dq}$ and $f(\alpha)=q \alpha(q) - \tau(q)$.
The $f(\alpha)$ vs $\alpha$ spectra of the STI and SI cases are plotted in 
Figs \ref{falph}(a) and \ref{falph}(b) respectively (where the 
 SI regime has been analysed omitting the gaps in the spectrum).
It is clear that STI of the DP class   
shows $f-\alpha$ behaviour  distinct from the SI class. The SI curves are more asymmetric and peak at higher values of $\alpha$. It is also interesting to note that STI of the DP class at distinct parameter values collapse quite closely on the same $f-\alpha$ curve for positive $q$ (since $\frac{df}{d\alpha}=q$, this is the part of the curve with positive slope), but separate out for negative $q-s$, whereas the data for 
the SI case does not fall on  the same curve for either regime.

\newpage

\pagebreak
\newpage

\vspace{.2in}

Fig. \ref{falph}(c) shows the comparison between the $f-\alpha$ spectrum of the SI distribution with gaps excluded (dotted line), and that where the gaps are included (diamonds, plotted for positive $q-s$, only). The contribution of the gaps can be very clearly seen. We also note that cross-over 
effects can be seen in the dynamical characterisers as well, and the 
distributions cross over from those characteristic of DP behaviour, to
those
characteristic of non-DP behaviour.  

Thus, the absence or presence of the gaps in the eigenvalue spectrum constitutes the primary
signature of DP and non-DP behaviour in this system. The leading
signature of DP or non-DP behaviour in the multifractal spectrum is the
divergence of the $\tau$ versus $q$ curve for negative $q-s$, as well as
corresponding behaviour in the $f-\alpha$ spectrum. However,
the secondary signatures of DP versus non-DP behaviour can be found even
when the $\tau$ versus $q$ for SI is obtained excluding the gaps which 
contribute to the divergence in the
curvature of the $\tau$ versus $q$ curve. 
Thus, the distribution of eigenvalues of the stability matrix and the multifractal spectrum 
of the distribution constitute the dynamic characterisers of DP and non
DP behaviour.

\section{Discussion}

To summarise, the phase diagram of the coupled sine circle map shows regimes of
spatiotemporal intermittency of different types. The regimes of
spatiotemporal intermittency with synchronized laminar states and
turbulent bursts are characterised by a complete set of DP exponents.
Regimes of spatial intermittency, where the bursts have regular temporal 
behaviour, do not show infective behaviour, and do not belong to the DP
class. Thus the same model can show DP and non-DP behaviour in different
regions of the parameter space. The signature of the DP and non-DP
behaviour can be seen in the dynamical characterisers of the system,
viz. the distribution of eigenvalues and the multifractal spectrum 
of this distribution. Gaps in the eigenvalue spectrum are characteristic
of spatial intermittency, i.e. of spatially localised, temporally
regular or quasi-regular bursts with associated non-DP exponents. The
eigenvalue spectrum is continuous for regimes of regular
spatiotemporal intermittency with spreading bursts and characteristic
DP exponents.  
The model also shows spatiotemporally  intermittent regimes with other 
types of laminar and burst states. The scaling behaviour in these
regimes, and the identification of their universality classes is being
pursued further. In order to gain insight into the way in which
correlations build up in this system, it may be useful to set up probabilistic cellular
automata which exhibit similar regimes and to examine their associated 
spin Hamiltonians \cite{Domany}. 
The comparison of the scaling exponents seen in this model with those
seen in absorbing phase transitions which do not belong to the DP class
\cite{chatepmpm,chatepmpm1,chatejk,hinrich1,odor1}
is
also of interest.
We hope to examine some of these questions in future
work.

\section{Acknowledgement}

NG thanks BRNS, India for partial support.
ZJ thanks CSIR for financial support.

\appendix
\section{ The definition of dp exponents }

The DP transition is characterised by a set of static and dynamic
critical exponents associated with various quantities of physical interest. 

\subsection{Static exponents}
\begin{enumerate}
\item[(i)] We first consider  the escape time $\tau(\Omega,\epsilon,L)$, which is defined as the time taken for the system starting from random initial conditions to relax to a completely laminar state. It is expected from finite-size scaling arguments that $\tau$ varies with the system size $L$ such that
\begin{displaymath}
\tau(\Omega,\epsilon)=\left\{ 
\begin{array}{ll}
\log L & \textrm{Laminar phase}\\
L^z & \textrm{Critical phase}\\
\exp L^c &\textrm{Turbulent phase}
\end{array}
\right.
\end{displaymath}
Hence, at the critical value of the coupling strength, $\epsilon_c$, the escape time $\tau$ shows a power law behaviour, $z$ being the associated exponent. 
\item[(ii)] The order parameter $m(t)$, associated with this transition is defined as the fraction of turbulent sites in the lattice at time $t$. At $\epsilon_c$, the order parameter scales as 
\beq
m\sim(\epsilon - \epsilon_c)^\beta, ~~~ \epsilon\rightarrow\epsilon^+
\eeq
At $t<< \tau$, $m(t)$ scales with $t$ as $m(\epsilon_c,t)\sim t^{-\beta/{\nu z}}$, where $\nu$ is exponent associated with the spatial correlation length.

The exponent $\nu$ is obtained by using the scaling relation
\beq
\tau(L,\epsilon_c)\sim \phi^z f(L/\phi)
\eeq
where, $\phi$ is the correlation length which diverges as $\phi\sim\delta^{-\nu}$ and $\delta$ is given by $(\epsilon$ - $\epsilon_c)$. Hence, $\nu$ is adjusted until the scaled variables $L\delta^{\nu}$ and $\tau\delta^{\nu z}$ collapse onto a single curve.

\item[(iii)] The correlation function in space is defined as 
\beq
C_j(t)=\frac{1}{L}\sum_{i=1}^L <x_i^t x_{i+j}^t> -< x_i^t>^2
\eeq
At $\epsilon_c$, $C_j(t)$ scales as $C_j(t)\sim j^{1-\eta'}$.\\

\item[(iv)] The distribution of laminar lengths, $P(l)$ is an important characteriser of the universality class \cite{Chate}. The laminar lengths, $l$ are defined as the number of laminar sites between two turbulent sites. At criticality, the laminar length distribution shows a power-law behaviour of the form
\beq
P(l)\sim l^{-\zeta}
\eeq
$\zeta$ is the associated exponent, $\zeta_{DP}$ being 1.67.
Another characteriser is the distribution of the laminar lengths which are $\geq l$ \cite{Grassberger}.This distribution shows a power-law behaviour of the form
\beq
\mathcal{P}(l)\sim l^{-\zeta'}
\eeq 

\end{enumerate}

\subsection{The dynamical exponents}

To extract the dynamical exponents, two turbulent seeds are placed in an absorbing lattice and the spreading of the turbulence in the lattice is studied. The quantities associated with critical exponents at $\epsilon_c$ are 
\begin{enumerate}
\item[(i)] the number of active sites, $N(t)$ at time $t$, which scales as $N(t)\sim t^{\eta}$
\item[(ii)] the survival probability, $P(t)$ defined as the fraction of initial conditions which show a non-zero number of active sites at time $t$. This scales as $P(t)\sim t^{-\delta}$, and 
\item[(iii)] the radius of gyration $R^2(t)$, which is defined as the mean squared deviation of the position of active sites from the original sites of turbulent activity. This scales as $R^2(t)\sim t^{z_s}$.
\end{enumerate}

\newpage


\begin{thebibliography}{99}
\bibitem{FN} The eigenvalue spectrum shown in Fig. \ref{nhgm}(a) shows that a tangent-period doubling bifurcation has occurred since the eigenvalues have crossed both $+1$ and $-1$. While we see this behaviour for some points where DP is observed, the more usual behaviour is where all the eigenvalues are positive and several eigenvalues have crossed $+1$, indicating that tangent-tangent bifurcations have occurred.
\end{thebibliography}
\end{document}